\documentclass[10pt,pdftex,a4paper,twocolumn]{article}
\usepackage[a4paper, total={7.2in, 10in}]{geometry}

\usepackage{meta}

\usepackage{mathtools}
\usepackage{amssymb}
\usepackage{amsmath}
\usepackage{amsfonts}
\usepackage{amsthm}
\usepackage{stmaryrd}

\usepackage[utf8x]{inputenc}
\usepackage{caption}
\usepackage{subcaption}   %
\usepackage{rotating}
\usepackage{ppprog}       %
\usepackage{shorthand}    %
\usepackage{veralgs}      %
\usepackage[OT1]{fontenc}
\usepackage{paralist}
\usepackage{notation}
\usepackage{xcolor}
\usepackage{mathtools}
\usepackage{framed}
\usepackage{accents}
\usepackage{lstlinebgrd}
\usepackage{tabularx}

\colorlet{shadecolor}{yellow!20}
\usepackage{tikz}
\usetikzlibrary{arrows,backgrounds,calc,decorations.pathmorphing,fit,shapes.geometric,matrix,shapes,patterns,positioning}
\usepackage{transactiondiagram}

\usepackage{enumerate}

\usepackage{codeconfig-bw}

\begin{document}

\title{\mytitle}

\author{\myauthors
\\\myemailgroup
\\\myinstitute, \myuniversity
\thanks{\mygrant}
}

\date{}

\maketitle

\begin{abstract}
\noindent Distributed Transactional Memory (DTM) is an emerging approach to distributed
synchronization based on the application of the transaction abstraction to
distributed computation. DTM comes in several system models, but the control
flow model (CF) is particularly powerful, since it allows transactions to
delegate computation to remote nodes as well as access shared data.  However,
there are no existing CF DTM systems that perform on par with
state-of-the-art systems operating in other models.  Hence, we introduce a CF
DTM synchronization algorithm, OptSVA-CF. It supports fine-grained pessimistic concurrency control, so it avoids
aborts, and thus avoids problems with irrevocable operations. Furthermore, it
uses early release and asynchrony to parallelize concurrent transactions to a
high degree, while retaining strong safety properties.  We implement it as
Atomic RMI 2, in effect producing a CF DTM system that, as our evaluation
shows, can outperform a state-of-the-art non-CF DTM such as HyFlow2.

\end{abstract}

\providecommand{\keywords}[1]{\noindent\textbf{Index terms---}#1}

\def\IEEEPARstart#1#2{#1#2}
\section{Introduction}
\label{sec:introduction}

\IEEEPARstart{W}{rangling} with concurrency in distributed computing is
difficult. Yet, in the era of cloud computing, when anything from simple text
editing to big data storage are delegated to remote services, this task is nigh
unavoidable.
Since application programmers have enough to worry about without delving into
the details of distributed computing and networking, these details are
comprehensively abstracted away and hidden within opaque libraries (e.g., Netty,
JGroups, Java Message Service, or the Java Remote Method Invocation mechanism)
to the point where such a programmer rarely resorts to directly using low-level
mechanisms like sockets.
Similarly, the problem of keeping concurrent execution correct should be so
hidden away under an abstraction, rather than expecting the programmer to do it
manually using primitives like locks, semaphores, or barriers and avoid their
various pitfalls like deadlocks, data races, or priority inversion.

Enter: \emph{distributed transactional memory} (DTM) \cite{BAC08,KAJ+08,CRCR09,SR11,SR12,TRP13,ADFK14}. 
DTM is an approach
modeled on database transactions and transactional memory known from
(single-host) multiprocessor systems \cite{HME93,ST95}. In this approach, the
programmer simply annotates which areas of code are transactions.  The DTM
system transparently ensures that code executed within transactions accesses
shared remote resources in a safe and consistent manner.

The thing that most starkly differentiates DTM from its database
predecessors is that apart from executing read and write operations on shared
resources (aka shared objects), these can provide interfaces for other or
different operations. Classically, this can be an operation like increment,
which both reads and writes the state of a shared object atomically, or stack
operations like push and pop. These can also be more complex,
computation-intensive, programmer-specified operations that execute just about
any code, and can include code with side effects.
By extension (and in contrast to non-distributed TM) it matters which processor
or which network node executes this code.
Therefore, it makes a difference what model of execution an implementation
uses. 
The \emph{data flow} (DF) model entails shared objects being migrated to
the client that uses them (while maintaining only a single copy of the object
in the system).  In such a case the computation and side effects will be
performed on whatever host the object is migrated to for executing an
operation. In the \emph{control flow} (CF) model, shared objects are bound to
individual hosts and do not migrate, so the execution of their operations is
performed always on the object's "home" host.
Both models have their advantages and disadvantages, 
but a unique feature of CF is that it allows to delegate computation to remote
hosts.
This allows client transactions to "borrow" computational power from remote
resource servers. In effect shared resources can act as both shared memory 
and web services. This provides greater flexibility in designing and implementing
distributed systems.

However, to the best of the authors' knowledge, there are currently no
well-performing CF DTM systems, and certainly none that would match a top notch DF system like HyFlow2
\cite{TRP13} which implements the \emph{Transaction Forwarding Algorithm} (\emph{TFA}) \cite{SR12}
for concurrency control. Its predecessor, HyFlow \cite{SR11} implements additional
algorithms, including DTL2 (a distributed variant of TL2 \cite{DSS06}) and has
CF support. However, as the authors show, HyFlow2 outperforms HyFlow by a
significant margin, to the point of obsoleting it.
The authors suppose, however, that there is no intrinsic problem stopping a CF
system from being equally efficient as a DF system.

In our previous research we introduced Atomic RMI \cite{SW15-ijpp}, a CR DTM
system that implements the \emph{Supremum Versioning Algorithm} (SVA), which employs
\emph{pessimistic} concurrency control. 
Typically, TM systems employ the \emph{optimistic} approach, 
where, generally speaking, a client transaction executes regardless of
other transactions running in parallel and performs validation only when it
finishes executing (at commit time). If two transactions try to access the same
shared object, and one of them writes to it, they \emph{conflict} and one of the transactions
aborts and restarts (in an optimized TM, this occurs as soon as possible, to waste
least work). When a transaction aborts, it should not change
the system state, so aborting transactions must revert the objects they
modified. Alternatively, they work on local copies and merge
them with the original object only on a successful commit. 
Pessimistic concurrency control is a different approach, known in database
transactions (e.g., two-phased locking)
and brought to TM in \cite{MS12,ADFK14} and our earlier work
\cite{Woj05b,Woj07}. It involves transactions waiting until they have
permission to access shared objects. In effect, potentially conflicting
operations are postponed until they no longer conflict. Thus transactions, for
the most part, avoid forced aborts, and thus, they also naturally avoid the
problems stemming from irrevocable operations.
The advantage of using pessimistic concurrency control over the optimistic
approach in a CF DTM system is that it does not require as stringent
assumptions with respect to what code can be part of transactions
or operations executed on shared resources.  Optimistic concurrency
control assumes implicitly that local operations executed within a
transaction (i.e., those not executed on shared objects directly) do not need to
be compensated for in the event of an abort, and can be safely re-executed.
This is not the case for a class of \emph{irrevocable operations}, which include acquiring
and releasing locks, system I/O, sending and consuming network messages. Such
operations are more likely to occur as the complexity of the code of
transactions increases, as can potentially happen with applications using CF to
delegate complex tasks to remote resources. 
The problem was mitigated in non-distributed TM by using special irrevocable
transactions that preempt other transactions and cannot abort (but only run
one-at-a-time) \cite{WSA08}, or providing multiple versions of objects that
transactions view for reads \cite{AH11,PFK10}. In other cases, irrevocable
operations are simply forbidden in transactions (e.g., in Haskell
\cite{HMJH04}).
Meanwhile, since pessimistic TM does not depend on aborts to maintain
consistency, the problem of irrevocable operations is mitigated for free.

The perceived problem of pessimistic TMs is that they struggle with performance
(see e.g., \cite{MS12}), but the main feature of SVA is the use of early release
to confer a performance improvement. Specifically, it uses \emph{a priori}
knowledge to detect when a transaction will certainly perform no further
operations on a particular shared object, and, in such a case, allows other
transactions to then access that object, rather than having them wait until the
first transaction commits.
However, since SVA is agnostic with respect to whether operations on shared
objects modify state or are read-only, it often perceives potential conflicts,
where those conflicts will not occur \emph{de facto}, and thus limits
parallelism more than strictly necessary. For this reason, Atomic RMI performs
similarly to HyFlow (with DTL2) and therefore is significantly outperformed by
HyFlow2 (with TFA).

In our most recent research \cite{WS16-arxiv}, we introduced a new pessimistic
TM concurrency control algorithm called \emph{Optimized SVA} (\emph{OptSVA}),
which introduces a number of optimizations to the basic \emph{modus operandi}
of its predecessor, mainly: heavy use of buffering rather than in-place
modifications, read operation parallelization, early release of shared objects
on last write instead of last operation of any kind, and asynchrony which
allows transactions to delegate some tasks that require waiting to separate
thread and proceed with other computation in the mean time. These
optimizations reduce the amount of scenarios where one transaction has to wait
for another and in this way improve the transactional throughput. 

However, OptSVA operates in a system model more typical for non-distributed
TMs, where shared objects are simple variables that can be either read from or
written to and cannot be trivially lifted to a general CF model.
Then, for instance, if the first operation on a variable in the transaction is
a write, an OptSVA transaction does not need to synchronize the state of the
variable with other transactions, but instead can simply store the value it
writes to a buffer and proceed. Furthermore, all subsequent reads are
local---they only depend on the value written by the current transaction, so
there is no need to synchronize before executing those reads either.  On the
other hand, the CF model allows operations to execute more complex code on the
shared object, and it may be that, for example, a write only modifies some
field {\tt a} of the object, but a subsequent read accesses its field {\tt b}.
In that case OptSVA would need to synchronize the state of the
object with other transactions before the read operation is allowed to execute,
but it does not supply a mechanism for that, so it is not directly useful in
the CF model.

The first contribution of the paper is to lift the OptSVA algorithm to the CF
system model, by re-designing it with complex programmer-defined shared objects
in mind. In this way we introduce OptSVA-CF, a more general variant of OptSVA,
that nevertheless is similarly capable of highly parallel execution of
concurrent transactions and that shares its predecessor's strong safety guarantees.

The second contribution of this paper is \atomicrmiii{}, an
implementation of OptSVA-CF that builds on Atomic RMI.
As with Atomic RMI, \atomicrmiii{} provides a simple-to-use API that allows
programmers to implement consistent transactions as simply as if using much
simpler mechanisms, such as distributed coarse-grained locking. However, the
application of OptSVA-CF allows for the execution of these transactions to be
highly parallelized.
We demonstrate this in a comprehensive evaluation of \atomicrmiii{}, showing
that it produces a significant efficiency increase over its predecessor as well
as various lock-based distributed concurrency control solutions. In addition,
we show that \atomicrmiii{} performs better than, or comparably to HyFlow2
(depending on contention and operation length), but does so while avoiding
aborting transactions altogether, thus allowing the use of irrevocable operations.

The paper is structured as follows. 
In \rsec{sec:optsva-cf}, we introduce the OptSVA-CF and explain how it works
and how it differs from its non-CF predecessor.
Then, in \rsec{sec:architecture} we proceed to discuss the implementation
details of OptSVA-CF within the framework of \atomicrmiii{}, including the
transactional API and the architecture.
In \rsec{sec:evaluation} we evaluate \atomicrmiii{} comparing it against Atomic
RMI---its predecessor, HyFlow2---a state-of-the-art DF DTM, and a number of
typical lock-based distributed synchronization solutions. Then, in
\rsec{sec:rw} we discuss related work, and conclude in
\rsec{sec:conclusions}.

\section{OptSVA-CF}
\label{sec:optsva-cf}

OptSVA-CF is a new pessimistic DTM concurrency control algorithm operating in
the CF model.
In this section we first describe the mechanisms the algorithm employs. We
follow by providing a summary of OptSVA-CF. 

\subsection{Concurrency Control through Versioning}
\label{sec:supremum-versioning}

In versioning algorithms, when each transaction starts, it is assigned a
\emph{private version} for each object it plans on accessing throughout its
lifetime. We denote transaction $\tr_i$'s private version for shared object
$\objx$ as $\pv{\objx}{i}$. The private versions are assigned from a sequence
of consecutive positive integers in such a way that:
\begin{enumerate}[a)] 
    \item %
        no two transactions have the same private version for any shared
        object, 
    \item %
        if transaction $\tr_i$ started before $\tr_j$ and they both access
        $\objx$, then $\pv{\objx}{i} < \pv{\objx}{j}$, 
    \item %
        given two transactions $\tr_i$ and $\tr_j$, if $\pv{\objx}{i} <
        \pv{\objx}{j}$ then for any shared object $\objy$ that both
        transactions plan to access, $\pv{\objy}{i} < \pv{\objy}{j}$, and 
    \item %
        if $\tr_i$ started before $\tr_j$ and no other transaction started in
        between the two, and both plan to access $\objx$, then they have
        consecutive private versions for $\objx$, i.e.,  $\pv{\objx}{i} =
        \pv{\objx}{j} - 1$.
\end{enumerate}

The versioning mechanism uses private versions to maintain order when accessing
shared objects via objects' \emph{local versions}.  That is, each shared object
$\objx$ has its own \emph{local version} counter, denoted $\lv{\objx}$,
which
is always equal to the private version of such
transaction $\tr_j$ that most recently finished using the object:
$\tr_j$ committed, aborted, or determined that it would no longer
need the object and released it early (see below). When $\tr_j$ does any of
those things, it writes its own private version to the local version
counter. Once $\tr_j$ releases the object in such a way it follows that some
other transaction can safely start calling methods on the object. 
We determine which transaction gets to access
the object next by simply selecting the transaction with the next consecutive
private version, i.e., such $\tr_i$ whose $\pv{\objx}{i} - 1 = \pv{\objx}{j}$.
Thus, invariably $\objx$ can be accessed by such $\tr_i$ for which
$\pv{\objx}{i} - 1 = \lv{\objx}$, and no other transaction. 
Hence, if some transaction $\tr_i$ wants to access $\objx$, then it may do so
if $\pv{\objx}{i} - 1 = \lv{\objx}$. We call this condition the \emph{access
condition}. On the other hand, if $\tr_i$ wants to access $\objx$ and the
access condition is not satisfied, then it waits until it is satisfied.

\def\magicadjust{\hspace{-.33cm}}
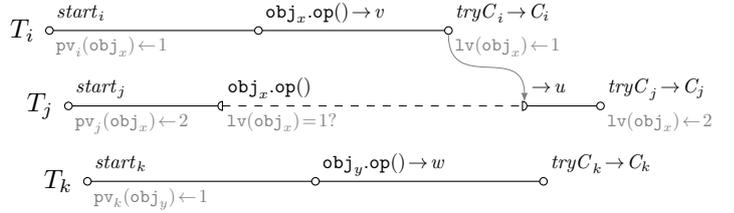
\begin{figure}[t]
\magicadjust
\begin{tikzpicture}
     \draw
           (0,2.5)      node[tid]       {$\trobj_i$}
                        node[aop]       {$\init_i$} %
                        node[dot]       {}
                        node[not]       {$\pv{\objx}{i}\!\gets\!1$}

      -- ++(2.75,0)     node[aop]       {$\objx.{\tt op()}\!\to\!\val$} 
                        node[dot]       {}

      -- ++(2.5,0)      node[aop]       {$\tryC_i\!\to\!\co_i$}
                        node[dot] (ci)  {}         
                        node[not]       {$\lv{\objx}\!\gets\!1$}
                        ;

     \draw
         (0.25,1.5)    node[tid]       {$\trobj_{j}$}
                        node[aop]       {$\init_{j}$} %
                        node[dot]       {} 
                        node[not]       {$\pv{\objx}{j}\!\gets\!2$}

      -- ++(2.00,0)     node[aop]       {$\objx.{\tt op()}$}
                        node[inv] (rx0) {}
                        node[not]       {$\lv{\objx}\!=\!1?$}
 
         ++(4.00,0)     node[aop]       {$\to\!\valu$}
                        node[res] (rx)  {}                       

      -- ++(1.,0)       node[aop]       {$\tryC_j\!\to\!\co_j$}
                        node[dot]       {}
                        node[not]       {$\lv{\objx}\!\gets\!2$}
                        ;

      \draw[hb] (ci) \squiggle (rx);
      \draw[wait] (rx0) -- (rx);

      \draw
           (0.5,0.5)      node[tid]       {$\trobj_k$}
                        node[aop]       {$\init_k$} %
                        node[dot]       {} 
                        node[not]       {$\pv{\objy}{k}\!\gets\!1$}

       -- ++(3.0,0)     node[aop]       {$\objy.{\tt op()}\!\to\!\valw$}
                        node[dot]       {}

      -- ++(3.00,0)     node[aop]       {$\tryC_k\!\to\!\co_k$}
                        node[dot]       {}         
                        node[not]       {}
                        ;
\end{tikzpicture}
\caption{\label{fig:access-control} Concurrency control via versioning.}
\end{figure}

An example of how this mechanism works is shown in \rfig{fig:access-control}.
The diagrams depict histories consisting of operations executed by transactions
on a time axis. Every line depicts the operations executed by a particular
transaction.
The symbol
\protect\tikz{
    \protect\draw[] (0,0) -- ++(0.25,0) node[dot] {} -- ++(0.25,0);
} 
denotes a complete operation execution.
The inscriptions above operation executions denote operations executed by the
transactions, e.g., $\objx.{\tt op()}\!\to\!\val$ denotes that an operation on
variable $\objx$ is executed by the transaction and returns some value $\val$, 
and $\tryC_i\!\to\!\co_i$ indicates that transaction $\tr_i$ attempts
to commit and succeeds because it returns $\co_i$.
On the other hand, the symbol 
\protect\tikz{
    \protect\draw[] (0,0) -- ++(0.1,0) node[inv] (a) {} 
                             ++(0.4,0) node[res] (b) {}
                          -- ++(0.1,0);
    \protect\draw[wait] (a) -- (b);
}
denotes an operation execution split into the invocation and the response
event to indicate waiting, or that the execution takes a long time.
In that case the inscription above is split between the events, e.g., a read
operation execution would show $\objx.{\tt op()}$ above the invocation, and
$\!\to\valu$ over the response.
If waiting is involved, the arrow 
\protect\tikz{
    \protect\draw[hb] (0,0.2) .. controls +(270:.25) and +(90:0.25) .. (0.5,0.0);
} 
is used to emphasize a happens-before relation between two events.
Annotations below events emphasize the state of counters or actions performed 
within the concurrency control algorithm (used as necessary).

In \rfig{fig:access-control}, $\trobj_i$ and $\trobj_j$ attempt to access
shared variable $\objx$ at the same time. Transaction $\trobj_i$ starts first,
so  $\pv{\objx}{i} = 1$, and $\trobj_j$ starts second, so $\pv{\objx}{j} = 2$.
Since initially $\lv{\objx} = 0$, $\trobj_j$ is not able to pass the access
condition and execute an operation on $\objx$ when it tries to, so it waits. On
the other hand, $\trobj_i$ can pass the access condition $\pv{\objx}{i} - 1 =
\lv{\objx}$ and it executes an operation on $\objx$ without waiting. Once
$\trobj_i$ commits, it sets $\lv{\objx}$ to $1$, so  $\trobj_j$ then becomes
capable of passing the access condition and finishing executing its operation
on $\objx$.
In the mean time, transaction $\trobj_k$ can proceed to access $\objy$
completely in parallel.

\subsection{Early Release}

Versioning algorithms use an early release mechanism to execute concurrent
transactions partially in parallel. A transaction can release any shared object
early at any point during its lifetime. This indicates that the transaction no
longer has any need to use the object, so some other transaction may start
using it instead.
Early release can be effected 
automatically by the
versioning algorithm. 
In order to release a shared object automatically, a
transaction must have \emph{a priori} knowledge of \emph{suprema}: it must know
at most how many times it will attempt to access each object throughout its
execution. If a supremum is defined for some object $\objx$, as the transaction
executes, it tracks how many times it actually calls $\objx$'s methods, and if
the number of actual calls reaches the supremum, the object can be assumed to
be no longer needed and released early immediately following a call. If the
supremum is never reached, the object is released on transaction completion. The
supremum should never be lower than the actual number of accesses to an object,
but if it is reached and a transaction subsequently calls the object
nevertheless, the transaction is immediately aborted.
Suprema are optional, and they can be provided by the programmer, or derived by
a type checker \cite{Woj05b}, or through static analysis \cite{SW12}.

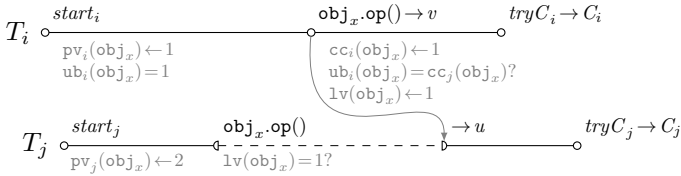
\begin{figure}[t]
\magicadjust
\begin{tikzpicture}
     \draw
           (0,2.5)      node[tid]       {$\trobj_i$}
                        node[aop]       {$\init_i$} %
                        node[dot]       {} 
                        node[not]       {
                                         \begin{multinote}
                                         $\pv{\objx}{i}\!\gets\!1$ \\
                                         $\ub{i}{\objx}\!=\!1$ \\
                                         \end{multinote}
                                         }

      -- ++(3.5,0)      node[aop]       {$\objx.{\tt op()}\!\to\!\valv$}
                        node[dot] (wx)  {}
                        node[not]       {
                                         \begin{multinote}
                                         $\cc{i}{\objx}\!\gets\!1$\\
                                         $\ub{i}{\objx}\!=\!\cc{j}{\objx}?$\\
                                         $\lv{\objx}\!\gets\!1$\\
                                         \end{multinote}
                                        }

      -- ++(2.5,0)      node[aop]       {$\tryC_i\!\to\!\co_i$}
                        node[dot] (ci)  {}         
                        ;

     \draw
         (0.25,1)       node[tid]       {$\trobj_{j}$}
                        node[aop]       {$\init_{j}$} %
                        node[dot]       {} 
                        node[not]       {$\pv{\objx}{j}\!\gets\!2$}

      -- ++(2.00,0)     node[aop]       {$\objx.{\tt op()}$}
                        node[inv] (rx0) {}
                        node[not]       {$\lv{\objx}\!=\!1?$}
 
         ++(3.00,0)     node[aop]       {$\to\!\valu$}
                        node[res] (rx)  {}                       

      -- ++(1.75,0)     node[aop]       {$\tryC_j\!\to\!\co_j$}
                        node[dot]       {}
                        node[not]       {}
                        ;

      \draw[hb] (wx) .. controls +(265:2) and +(90:1) .. (rx);
      \draw[wait] (rx0) -- (rx);

\end{tikzpicture}
\vspace{-.1cm}
\caption{\label{fig:early-release} Early release via upper bounds.}
\end{figure}

The early release mechanism is illustrated further in \rfig{fig:early-release}.
Here, transactions
$\trobj_i$ and $\trobj_j$ both try to access $\objx$. Like in
\rfig{fig:access-control}, since $\trobj_i$'s private version for $\objx$ is
lower than $\trobj_j$'s, the former manages to access $\objx$ first, and
$\trobj_j$ waits until $\trobj_i$ is released. Unlike in
\rfig{fig:access-control}, $\trobj_i$ has upper bound information:
it knows that it will
execute at most one operation on $\objx$ ($\ub{i}{\objx} = 1$).
The number of operations executed on $\objx$ is tracked using counter $\cc{i}{\objx}$
So, $\trobj_i$ releases $\objx$ immediately after it accesses $\objx$, rather than
waiting to do so until commit. In effect, $\trobj_j$ can access $\objx$
earlier.

\subsection{Aborts}

The concurrency control algorithm is pessimistic in nature, and does not need
to abort any transaction to ensure correct execution. Even so,  a way to
manually abort transactions is provided to the programmer, since this makes it
easier to cancel a transaction mid-execution without having to manually scrub
its effects. More importantly, it makes the implementation of such features as
fault tolerance possible.
However, in conjunction with early release, aborts also introduce a possibility
of transactions reading inconsistent state. We can imagine a situation where a
transaction releases an object early, and another transaction starts accessing
it, but the first transaction eventually aborts. In that case the second
transaction cannot be allowed to commit, since it accessed data that was later
invalidated.

In versioning algorithms that allow aborts the scenario is resolved by
maintaining the order in which transactions execute a commit or abort on each
variable, by analogy to accessing objects.
Each shared object has a \emph{local terminal version} counter $\ltv{\objx}$
which holds the private version of the transaction that either committed or
aborted last. If some transaction $\tr_j$ wishes to commit or abort it may only
do so if $\pv{\objx}{j} - 1 < \ltv{\objx}$, which we call the commit condition,
and otherwise it must wait. In effect, if transaction $\tr_i$ accesses $\objx$
before $\tr_j$, it is ensured that $\tr_i$ commits before $\tr_j$.
It means that the algorithm has a chance to forcibly abort a transaction.

Any transaction that aborts writes, marks each object in its access set as an
invalid instance, and reverts each object's
state. 
Each transaction checks whether the object is valid before accessing it
or executing commit- or abort-related activities on it. If the object is
invalid, then the transaction is forced to abort instead of executing whatever
operation it was supposed to execute. Since the commit operations are ordered
according to private versions, a transaction will not abort unless it is
impossible for any of the objects it accessed to be invalidated.

\begin{figure}[t]
\magicadjust
\begin{tikzpicture}
     \draw
           (0,2.5)      node[tid]       {$\trobj_i$}
                        node[aop]       {$\init_i$} %
                        node[dot]       {} 
                        node[not]       {
                                            $\pv{\objx}{i}\!\gets\!1$
                                        }

      -- ++(2.25,0)     node[aop]       {$\objx.{\tt op()}\!\to\!\valv$}
                        node[dot] (wx)  {}
                        node[not]       {$\lv{\objx}\!\gets\!1$}

      -- ++(2.5,0)      node[aop]       {$\tryA_i\!\to\!\ab_i$}
                        node[dot] (ci)  {}
                        node[not]       {
                                            \begin{multinote}
                                            $\ltv{\objx}\!\gets\!1$   \\
                                            ${\tt invalidate}\;\objx$ \\
                                            \end{multinote}
                                        }
                        ;

     \draw
         (0.25,1)       node[tid]       {$\trobj_{j}$}
                        node[aop]       {$\init_{j}$} %
                        node[dot]       {} 
                        node[not]       {
                                            $\pv{\objx}{j}\!\gets\!2$
                                        }

      -- ++(1.75,0)     node[aop]       {$\objx.{\tt op()}$}
                        node[inv] (rx0) {}
                        node[not]       {${\tt is}\;\objx\;{\tt valid?}$}
 
         ++(2,0)        node[aop]       {$\to\!\valu$}
                        node[res] (rx)  {}                       

      -- ++(1.,0)       node[aop]       {$\tryC_j$}
                        node[inv] (cj0) {}
                        node[not]       {$\ltv{\objx}\!=\!1?$} 

         ++(2,0)        node[aop]       {$\to\!\ab_j$}
                        node[res] (cj)  {}
                        node[not]       {${\tt is}\;\objx\!{\tt valid?}$}
                        ;

      \draw[hb] (wx) .. controls +(265:1.5) and +(90:1) .. (rx);
      \draw[hb] (ci) .. controls +(265:1.75) and +(90:1) .. (cj);
      \draw[wait] (rx0) -- (rx);
      \draw[wait] (cj0) -- (cj);

\end{tikzpicture}
\caption{\label{fig:commit-order} Cascading abort via versioning.}
\end{figure}
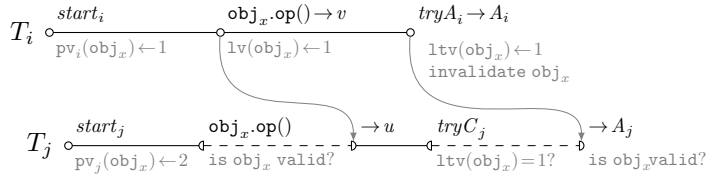

An example of how this mechanism affects execution is shown in
\rfig{fig:commit-order}.  Here, $\trobj_i$, $\trobj_j$ both access $\objx$ and
they respectively get the values of $\pvf$ for $\objx$ of $1$ and $2$.
Transaction $\trobj_i$ accesses $\objx$ first and releases it early, setting
$\lv{\objx}$ to $1$. This allows $\trobj_j$ to pass the access condition and
access $\objx$. Transaction $\trobj_j$ subsequently attempts to commit.
However, in order to commit $\trobj_j$ must pass the termination condition
$\pv{\objx}{j} - 1 = \ltv{\objx}$, which will not be satisfied until $\trobj_i$
sets $\ltv{\objx}$ to its own $\pv{\objx}{i}$. Hence $\trobj_j$ can only
complete to commit until $\trobj_i$ commits or aborts. When $\trobj_i$
eventually aborts it invalidates $\obj$. When $\trobj_i$ attempts to commit, it
cannot do so, because $\objx$ is invalid, so it finally aborts.

\subsection{Irrevocable Operations}

The versioning mechanism is pessimistic: it delays 
operations rather then aborting transactions on conflict.
Transactions only abort if the abort operation is invoked manually by the
programmer, or as a result of a cascade.%
Further, cascading aborts start only due to a transaction being aborted manually.
Hence, if no transaction in the system manually aborts, no transaction ever
aborts. Then, it is to execute any irrevocable operations
within any transaction.

However, if any transaction manually aborts, it is possible that it
will force some other transaction into a cascade.
In order for transaction $\tr_i$ to participate in a cascading abort, a
preceding transaction $\tr_j$ must release an object early and then abort after
the $\tr_i$ executed an operation directly on that object.
These conditions rarely occur in practice, so cascading aborts are also infrequent.
However, they do introduce a chance of unsafe executions of irrevocable operations.

Hence, in order to exclude the possibility completely for transactions with
irrevocable operations, such transactions can be labelled \emph{irrevocable
transactions}. OptSVA-CF prevents irrevocable transactions from ever becoming
part of a cascade, by replacing all access condition checks with termination
condition checks. This means that irrevocable transactions never "accept"
objects released early. The drawback is that such transactions may wait longer
to access shared objects, but in return they never forcibly abort.

\subsection{Operations in the Complex Object Model}

Initially, versioning algorithms were conceived as operation-type agnostic,
which made them suitable for use with \emph{complex shared objects} using in
the CF model.
These objects have arbitrary interfaces, whose operations (methods) execute
arbitrary computations on state that can be composed of multiple discrete
variables (fields).
Such operations may not be limited to reading or writing the state of the
object, but may do both, or neither, or cause side-effects in the process.
Furthermore, each object may have a different interface.
It is therefore practical to treat such objects as black boxes 
with respect to their state and operations they execute.

Contrast this to simple objects, variables, used commonly in TM (see e.g.,
\cite{GK10}), where each object has a single read operation that reads the
state of the object, and a single write operation which supersedes the previous
state of the object with a new state. Both operations are simple, completely
transparent, and contain no side effects, which allows to better orchestrate
their execution.

An apt example 
is that
of operation \emph{locality}. 
According to \cite{GK10}, a \emph{local read} is a read that is preceded during
transaction execution by a write on the same shared variable---because it only depends on
state written by that write operation, so it does not depend on what other
transactions write. 
A \emph{local write} is a write that is followed by a write on the same
variable---because whatever that first write writes is superseded by the value
written by the second. 
Local operation executions do not impact the system outside of their
transaction.
Thus, buffering can be used to make them invisible to the outside world.
A local write modifies a transaction-local buffer, rather than the actual
object. This means that local writes \emph{de facto} do not operate on shared
objects, so they do not need to pass the access condition to be executed.
As we showed with OptSVA \cite{WS16-arxiv}, using such optimizations with this
model means that transactions execute more in parallel, and produce tighter
schedules as a result, which improves system throughput.

On the other hand, the simplification of complex objects to variables restricts
the flexibility of such a system model and limits its applicability in
distributed systems.
This is especially true in the CF model, where a complex object can be used not
only to store and retrieve data, but also to delegate more involved, possibly
long-running computations to a remote host. 
Once the arbitrary nature of
interfaces and operation semantics is removed, the latter is lost and the system model loses
its expressiveness.
Hence we introduce the distinction between read and write
operations in the complex object model 
with arbitrary interfaces 
by requiring
that each operation be classified as one of the following:
\begin{enumerate}[a) ]
    \item a \emph{read} operation is any operation that executes any code
        (including code with side effects) and may read the shared object's
        state and return a value, but during execution the state is never
        modified,  
    \item a \emph{write} operation is any operation that executes any code and
        may modify the state of the shared object, but the state is not read,
        nor modified,
    \item an \emph{update} operation is any operation that executes any code and
        may both modify and read the object's state and return a value.
\end{enumerate}
This classification allows us to mimic the optimizations used with
variable-like objects within complex object synchronization, but without
knowing the details of each operation's semantics.
We introduce the update operation, because we expect a
typical operation on a complex object to modify its state based
on the existing state of the object, hence to behave both like a read and write. 
However, such an operation is difficult to make invisible and parallelize.
On the other hand, "pure" writes, can be expected to be rare, but they
do not need to view the state to execute, so more optimizations apply to them.
Specifically, they
can also be made to execute on an "empty" buffer without prior synchronization.
Thus, we keep them apart from updates.
Note that the complex shared object may still contain composite state,
consisting of some number of independent variables, and read, write, and update
operations are not required to read and/or modify the state holistically. 
Whether or not a particular operation will only read state written locally or
whether it requires synchronization depends largely on how objects buffering 
is implemented.

\subsection{Buffering}

When creating buffers for variable-like objects, given the semantics of
the two available operations, it is simply a matter of copying a value from a
shared variable to some local variable. Such a buffer can also be locally written to
without knowledge of its state, since the new value of the variable supersedes
the old. Thus, local writes can simply write to uninitialized local variables.

Given the composite state of complex objects and arbitrary semantics of
operations, two types of buffers are needed. The first, a \emph{copy buffer},
is one that copies the entire state of a shared object, and can be used to both
locally read and modify the object. Such a buffer can be used to read a
released object or restore an object during abort. 
However, since the state of the original object is copied, in order to create a
copy buffer the transaction must check fulfill the access condition before
doing so. Such a copy buffer is not universal, since it cannot be used to
execute local writes without prior synchronization. 

Thus, we introduce a second buffer type.
A \emph{log buffer} is an object that maintains the interface of the
original shared object but none of its state. When a method is executed on the
object, the buffer logs the method and its parameters. The method may be
executed completely, assuming that it does not need any state other than local
data to do so. In that case, any changes the method does to the state are
tracked and stored. If this is impossible, the method will not execute, apart
from being logged.
The log buffer can be applied to the original object to update the state of the
latter. If some method was pre-executed before applying the log, its effects
are applied to the state of the original object. If a method was not previously
executed, it is executed on the original object at the time the log is being
applied.
Given the log buffer does not use the object's state, it can be used to execute
write operations without prior synchronization. Since write operations modify
the object's state without viewing it, write operations are always capable of
executing methods on the log buffer in place, and do not need to commute the
execution to the point when the buffer is applied.

Since the CF semantics require that computations are performed wherever the
shared object is located in the distributed system, either type of buffer
resides on the same host in a distributed system as the original object. 
Otherwise, not only would the assumptions of the CF model be violated, but
if the execution of operations caused any side effects, 
the side effects would be removed from the location of the original node.

\subsection{Asynchronous Buffering}

\begin{figure}[t]
\magicadjust
\begin{tikzpicture}
     \draw
           (0,2.5)      node[tid]       {$\trobj_i$}
                        node[aop]       {$\init_i$} %
                        node[dot]       {} 
                        node[not]       {$\pv{\objx}{i}\!\gets\!1$}

      -- ++(2.0,0)      node[aop]       {$\objx.{\tt write(1)}\!\to\!()$}
                        node[dot]  (wx) {}
                        node[not]       {$\lv{\obj}\!\gets\!1$}

      -- ++(3.0,0)      node[aop]       {$\tryC_i\!\to\!\co_i$}
                        node[dot]  (ci) {}
                        ;

     \draw
          (0.25,1.25)   node[tid]         {$\trobj_j$}
                        node[aop]         {$\init_j$} %
                        node[dot] (fork1) {} 
                        node[not]         {$\pv{\objx}{j}\!\gets\!2$}

      -- ++(2.25,0)     node            {}
      -- ++(.08,0)      node    (join1) {}

      -- ++(.25,0)      node[aop]       {$\objx.{\tt read()}\!\to\!1$}
                        node[dot]       {}

      -- ++(2.25,0)      node[aop]       {$\objx.{\tt read()}\!\to\!1$}
                        node[dot]       {}

      -- ++(2.,0)     node[aop]       {$\tryC_j\!\to\!\co_j$}
                        node[dot] (cj)  {}
                        ;

      \draw
           (.25,.6)     node[inv] (ro1) {} %
                        node[not]       {$\lv{\objx}\!=\!1?$}
 
          ++(2.25,0)    node {}
          ++(.08,0)     node[res] (ro2) {}
                        node[not]       {
                            \begin{multinote}
                            $\buf{j}{\objx}\!\gets\!\objx$ \\
                            $\lv{\objx}\!\gets\!2$ \\
                            \end{multinote}
                            }
                        ;
                       
      \draw[] (fork1) -- (ro1);
      \draw[] (join1.center) -- (ro2);
      \draw[wait] (ro1) -- (ro2);
      \draw[hb] (wx) \squiggle (join1.center);

     \draw
         (0.5,-.75)     node[tid]       {$\trobj_{k}$}
                        node[aop]       {$\init_{k}$} %
                        node[dot]       {} 
                        node[not]       {$\pv{\objx}{k}\!\gets\!3$}

      -- ++(2.0,0)      node[aop]       {$\objx.{\tt read()}$}
                        node[inv] (rx0) {}                     
                        node[not]       {$\lv{\objx}\!=\!2?$}

         ++(1.8,0)       node[aop]       {$\!\to\!1$}
                        node[res] (rx2) {} 
                       
      -- ++(.6,0)      node[aop]       {$\objx.{\tt write(2)}\!\to\!()$}
                        node[dot]        {}                     

       -- ++(2.5,0)     node[aop]       {$\tryC_k\!\to\!\co_k$}
                        node[dot] (ck)  {}
                        ;

      \draw[hb] (ro2) .. controls +(265:1.65) and +(90:1) .. (rx2);
      \draw[wait] (rx0) -- (rx2);
\end{tikzpicture}
\caption{\label{fig:read-only-opt} Asynchronous read-only buffering. 
}
\end{figure}
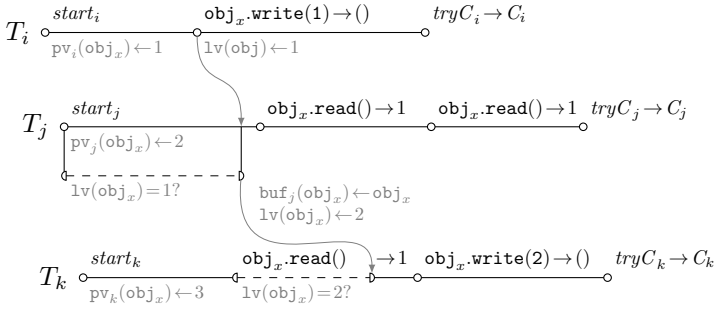
\begin{figure*}[t]
\begin{center}
\begin{tikzpicture}
     \draw
           (0,4)        node[tid]       {$\trobj_i$}
                        node[aop]       {$\init_i$} %
                        node[dot]       {} 
                        node[not]       {$\pv{\objx}{i}\!\gets\!1$}

      -- ++(2.5,0)      node[aop]       {$\objx{\tt .read()}\!\to\!0$}
                        node[dot]  (rx) {}
                        node[not]       {$\lv{\obj}\!=\!0?$}

      -- ++(3.,0)       node[aop]       {$\objx.{\tt write(1)}\!\to\!()$}
                        node[dot]  (wx) {}
                        node[not]       {$\lv{\obj}\!\gets\!1$}

      -- ++(3.,0)       node[aop]       {$\tryC_i\!\to\!\co_i$}
                        node[dot]  (ci) {}
                        ;

     \draw
           (0.25,2.25)  node[tid]       {$\trobj_j$}
                        node[aop]       {$\init_j$} %
                        node[dot]       {} 
                        node[not]       {
                                            \begin{multinote}
                                            $\wub{j}{\objx}\!=\!2$ \\
                                            $\pv{\objx}{j}\!\gets\!2$\\
                                            \end{multinote}
                                        }

       -- ++(2.25,0)    node[aop]       {$\objx.{\tt write(2)}\!\to\!()$}
                        node[dot] (wjx)      {}
                        node[not]       {
                            \begin{multinote}
                            $\log{j}{\objx}.{\tt write(2)}$ \\
                            $\wc{j}{\objx}\!\gets\!1$ \\
                            $\wub{j}{\objx}\!=\wc{j}{\objx}?$ \\
                            \end{multinote}
                        }
      
      -- ++(3.,0)       node[aop]         {$\objx{\tt .write(3)}\!\to\!()$}
                        node[dot] (rx0j){}
                        node[not]       {
                            \begin{multinote}
                            $\log{j}{\objx}.{\tt write(3)}$ \\
                            $\wc{j}{\objx}\!\gets\!2$ \\
                            $\wub{j}{\objx}\!=\wc{j}{\objx}?$ \\
                            \end{multinote}
                        }

      -- ++(3.,0)       node[aop]         {$\objy{\tt .update()}\!\to\!()$}
                        node[dot] {}

      -- ++(2.5,0)      node (join) {}

      -- ++(2.5,0)       node[aop] {$\objx.{\tt read()}\!\to\!2$}
                        node[dot] {}

      -- ++(2.25,0)      node[aop]       {$\tryC_j\!\to\!\co_j$}
                        node[dot] (cj) {}
                        ;

       \draw
            (5.5,1)   node[inv] (ro1) {} %
                         node[not]     {$\lv{\objx}\!=\!1?$}
  
           ++(5.5,0)    node[res] (ro2) {}
                         node[not]  {
                             \begin{multinote}
                             $\objx\!\gets\!\log{j}{\objx}$\\
                             $\buf{j}{\objx}\!\gets\!\objx$\\
                             $\lv{\objx}\!\gets\!2$\\
                             \end{multinote}
                         }
                        ;

      \draw[] (rx0j) -- (ro1);
      \draw[] (ro2) -- (join.center);
      \draw[wait] (ro1) -- (ro2);

      \draw[hb] (wx) 
        .. controls +(265:2) and +(90:1.5) .. 
        (join.center);

     \draw
          (3.5,-.75)    node[tid]       {$\trobj_{k}$}
                        node[aop]       {$\init_{k}$} %
                        node[dot]       {} 
                        node[not]       {$\pv{\objx}{k}\!\gets\!3$}

      -- ++(2.5,0)      node[aop]       {$\objx{\tt .read()}$}
                        node[inv] (krx0) {}
                        node[not]       {$\lv{\objx}\!=\!2?$}

         ++(6.5,0)      node[aop]       {$\to\!3$}
                        node[res] (krx) {}                       

      -- ++(.75,0)      node[aop]       {$\objx.{\tt write(4)}\!\to\!()$}
                        node[dot]       {}

      -- ++(2.75,0)     node[aop]       {$\tryC_k\!\to\!\co_k$}
                        node[dot]       {}
                        node[not]       {}
                        ;

      \draw[hb] (ro2) .. controls +(265:2) and +(90:0.75) .. (krx);
      \draw[wait] (krx0) -- (krx);

\end{tikzpicture}
\caption{\label{fig:initial-writes-opt} Asynchronous release on last write.
}
\end{center}
\end{figure*}
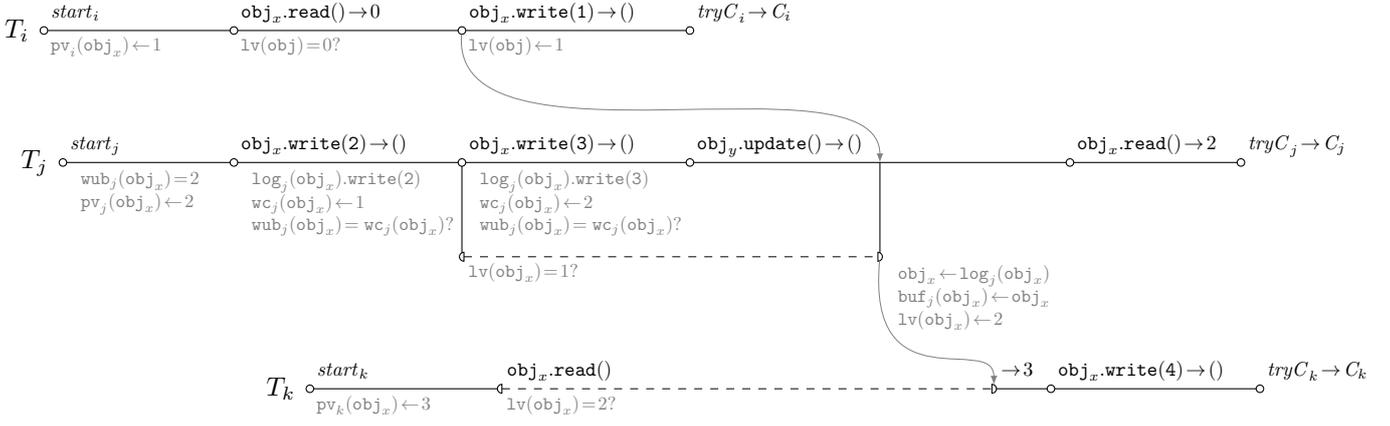

A special case occurs if a transaction only ever executes read operations on a
shared object (although it may execute writes and updates on other objects). We
will refer to such an object as a \emph{read-only object} with respect to this
transaction. In the case of such an object, synchronization needs to be done
when the first read is executed, but all subsequent reads
only need to use the buffer to execute. Hence, once the buffer is created,
the reads execute as if they were local and
the read-only object can be released.

Other transactions benefit from 
the object being released as soon as possible, and it is possible for a
read-only object to be released even before the first read operation occurs.
I.e., first write does not need to access the actual object either, as long as
the state of the object is buffered. The only condition that must be satisfied
to store the state of some object in the buffer is that it must pass the access
condition, but otherwise it can be done at any point in the transaction. Hence,
OptSVA-CF transactions attempt to buffer a read-only object as soon as they
retrieve private versions at start. But, since waiting at the access condition
may block the executing of operations that precede the first read on the
read-only object in the code of the transaction, the buffering procedure is
executed asynchronously: the transaction delegates it to a separate thread and
proceeds to execute other operations as normal. 
The separate thread waits until the access condition for the object is met,
following which the thread buffers, and immediately releases the object. Then,
all reads, including the first read, execute the operation on the buffer. In
effect, early release of read-only objects is potentially expedited.

We show an example of this in \rfig{fig:read-only-opt}. Here, transaction
$\tr_j$ treats $\objx$ as a read-only object, and tries buffering it as soon as
it starts. This is done in a separate thread (indicated by the line below)
which waits until the access condition is satisfied for $\objx$. Once $\tr_i$
releases $\objx$, the thread immediately buffers the object and releases it.
This allows transaction $\tr_k$ to begin accessing and modifying $\objx$, even
while $\tr_j$ executes two read operations in parallel using its buffer.
If the buffer were not used, $\tr_j$ would delay operations performed by
$\tr_k$.

Similar asynchrony is used in the case of a final modification of an object.
When a transaction executes its last write or update operation on some shared
object, 
the object is immediately buffered afterward and released. This allows all
following read operations to only use the buffer, and therefore be invisible to
outside transactions.
The final update can only be executed if the access condition is passed, since
it may need to view the state of the object to execute. However, a write may
execute using the log buffer instead and without synchronization, since it does
not view the state.
Then, in the specific case of a write operation that is the only write
operation on an object, or in case of a write operation preceded only by other
write operations on that object, the transaction may not have attempted to
satisfy the access condition yet. In such a case, the final write can be split
into a write that executes using the log buffer without synchronization,
and a procedure that subsequently updates the state of the actual object. This
procedure can only be executed if the access condition is passed, but it can
release the object immediately after it finishes updating the object's state.
The procedure is executed asynchronously with respect to the main body of the
transaction, since it has no impact on following operations---all following
operations on this object, if any, will be reads, and will read from the local
buffer. In this way, the last write avoids blocking the entire transaction to
wait for the access condition. In addition, the point at which the transaction
must wait for the access condition for this object can be delayed to any point
between the last write and the end of the transaction.

We illustrate this optimization further in \rfig{fig:initial-writes-opt}. Here, transaction
$\trobj_i$ can pass access condition for $\objx$ first and execute a read and a
write on $\objx$.
Nevertheless $\trobj_j$ performs operations on $\objx$ simultaneously.
First, $\tr_j$ executes a write, and can do so without waiting at the access
condition, since it works on the log buffer rather than directly on $\obj$.
Meanwhile $\tr_i$ can execute operations on the actual object.  Then $\tr_j$
executes another write operation, using the log buffer. Since this is the last
operation execution on $\obj$ in $\tr_j$ (which the transaction knows because
$\wub{j}{\objx} = \wc{j}{\objx}$), $\tr_j$ may write the changes from the log
buffer into the object. 
Hence, a separate thread starts at the end of the write operation and it starts
waiting at the access condition. When the access condition is satisfied, the
thread updates the state of $\objx$ using the log buffer and releases it, which
allows $\tr_k$ to start accessing $\objx$. The thread also creates a copy
buffer from the updated object, which is sufficient for future local reads to
use.
Note, however, that $\tr_j$ can immediately start doing other operations,
while the separate thread is still waiting at the access condition to $\objx$.
Hence $\tr_j$ can execute an update operation on $\objy$, which can be executed
regardless of the access condition to $\objx$. Then, $\tr_j$ can continues to
execute read operations on $\objx$ using the copy buffer, and does so in
parallel to $\tr_k$.
If buffers were not used, none of these operations could be executed in
parallel by several transactions. In addition, if a separate thread were not
used to synchronize and release $\objx$, but if instead this were done as part
of the last write, the operation on $\objy$ in $\tr_j$ would be significantly,
but needlessly delayed.

\subsection{Summary}

OptSVA-CF operates on the basis of the versioning mechanism, using
private, local, and local terminal version counters to ensure that accesses to
objects and commits are performed in the order defined by private versions. 
The individual operations 
are handled as follows:

\subsubsection{Start}
When an OptSVA-CF transaction $\tr_i$ starts it acquires a private version for each
shared object in its access set. If any of these objects are read-only, the
transaction also starts separate threads that clone the objects into copy
buffers $\buf{i}{\objx}$ and release them afterwards.

\subsubsection{Read}
Whenever transaction $\tr_i$ attempts to execute a read operation on some
object $\objx$, its behavior primarily depends on whether the object is
read-only, and whether it was released or not.
If the object is read-only, the read operation waits until the separate
read-only thread finishes buffering the object, and executes the read operation
on the buffer.

Otherwise, if the object was not previously accessed, then the transaction
checks if there were preceding reads or updates. If not, the transaction must
wait until the access condition to $\objx$ is satisfied and makes a checkpoint
by copying the state of the shared object to buffer $\stored{i}{\objx}$. Buffer
$\stored{i}{\objx}$ is a copy buffer like $\buf{i}{\objx}$, but it
is never modified and only used to restore the object in the event of abort.

In addition, if only preceding operations
were writes, then they were performed using the log buffer $\log{i}{\objx}$, so
the transaction applies the log buffer to $\objx$ before proceeding. 
Next, the transaction checks if any object was invalidated so far, and if so
forcibly aborts. If any object was invalidated at any point, the transaction is
doomed to abort eventually, so by checking for all the objects we force it to
do so as early as we can detect. In any case, the transaction then executes
the code of the read operation on $\objx$. If this is the last operation of any
kind of $\objx$, the transaction subsequently releases $\objx$.

If the object was previously released, the read waits until the thread
responsible for releasing the object is finished. Once this is the case, the
transaction executes the code of the read operation on $\objx$ using the copy
buffer $\buf{i}{\objx}$ (created at release).

\subsubsection{Update}
In the case of an update operation, the transaction checks whether any reads or
updates were executed on the same object before. If that is the case, the
transaction waits until the access condition is satisfied and makes a
checkpoint. In addition, the transaction will also apply the log buffer
$\log{i}{\objx}$ to $\objx$ if there were preceding writes (but no preceding
reads or updates). In any case, the transaction subsequently checks whether any
objects were invalidated, and aborts if that is the case. Afterwards, the code
of the operation is executed on $\objx$. If there are no further
updates or writes to be performed on $\objx$, the transaction makes a copy of
$\objx$ in $\buf{i}{\objx}$ and releases it.

\subsubsection{Write}
Pure write operations are executed in one of two ways, depending on whether
there were any read or update operations executed prior. This is because
updates and reads both wait on the access condition, meaning that then the
object can be operated on directly. Otherwise, the write can be performed using
a log buffer.
Specifically, if there were no preceding reads or updates, the transaction
simply executes the operation on the log buffer. If this is the final write and
there will also not be update operations on this object in the transaction, the
transaction then starts a thread, which will wait at the access condition and
subsequently: make a checkpoint to $\stored{i}{\objx}$, apply the log buffer
$\log{i}{\objx}$ to the original object $\objx$, copy the modified object to
the copy buffer $\buf{i}{\objx}$, and release $\objx$. Meanwhile, the
transaction's main thread proceeds.

If there were preceding reads or updates, the transaction operates
using the up-to-date object that is already under its control. Making a
checkpoint would be redundant, but the transaction checks whether any objects
were invalidated, and if so, aborts. Otherwise, it executes the code of the
operation on the object, and if this was the last write or update operation on
$\objx$, then $\objx$ is cloned to $\stored{i}{\objx}$ and released. The
last step is not done in a separate thread, since the transaction already has
access to $\objx$.

\subsubsection{Commit}
When the transaction commits it waits for extant threads to finish in the case
such threads are still running for read-only objects and objects that are being
released after last write.  Afterwards, the transaction waits until the commit
condition is satisfied for all objects in its access set. 
Then, if the transaction did not access a particular object at any time, it
makes a checkpoint. If it only ever executed writes on an object, the
transaction applies the log buffer to the object. If the object was not
released, the transaction releases it.
Afterward, the transaction checks whether any object was invalidated, and
aborts if that is the case.
Otherwise, the transaction updates the local terminal versions of all objects
and finishes execution.
No further actions may be performed by the transaction after the commit finishes
executing.

\subsubsection{Abort}
When the transaction aborts, just like with commit, it waits for the
appropriate threads to finish, and for the commit condition to be satisfied.
Then, each object in the transaction's access set is restored from
$\stored{i}{\objx}$, unless some other transaction that previously aborted
already restored it to an older version beforehand. Then, the transaction updates the
local terminal versions of all objects and finishes execution. No further
actions may be performed by the transaction after the abort finishes executing.

\subsection{Consequences of Model Generalization}

OptSVA-CF is based on the optimizations introduced in OptSVA, but applies them
to a different, more universal system model. The complex object model is more
general, since a variable-like object can be implemented as
a reference cell, 
a complex object with one field, a read operation returning its
value, and a write operation setting the old value to a new one.

Given such a specification, OptSVA-CF will execute the same way as OptSVA with
one exception. Given a transaction that executes a write operation on some
object $\objx$ followed by a read operation on $\objx$, OptSVA-CF will execute
the write without synchronization, but must synchronize before the read
executes. This is because the changes in the log buffer must be applied to the
actual object and copy buffer before the read proceeds. Hence, the read might
be forced to wait until the access condition for $\objx$ is satisfied. 
In contrast, OptSVA will allow the read to proceed without synchronization,
since it is recognizable as a local operation, and therefore completely
dependent on the preceding write. 

In effect, given reference cells, there are certain executions that will be
allowed by OptSVA that are not allowed by OptSVA-CF. Since these OptSVA
executions are tighter than their equivalent executions in OptSVA-CF, OptSVA
admits a higher level of parallelism. Therefore, OptSVA-CF trades generality
for performance.

\subsection{Properties}

We briefly demonstrate the safety, liveness, and progress guarantees of OptSVA-CF.

\subsubsection{Safety}

In \cite{WS16-arxiv} we demonstrate that OptSVA is last-use opaque. The
requirements in synchronizing complex objects mean that if OptSVA-CF is used to
model the system model of OptSVA, OptSVA-CF allows a subset of executions
allowed by OptSVA. Hence, OptSVA-CF is also last-use opaque
\cite{SW14-disc,SW15-arxiv}. This implies that OptSVA-CF is
serializable, recoverable, preserves real-time order, and does not allow
overwriting a value once an object is released.

Furthermore, if the manual abort operation is never used within a given system,
OptSVA-CF never causes any transaction to abort, meaning that such OptSVA-CF
executions are indistinguishable from opaque \cite{GK10} (as shown in
\cite{SW15-fcds}, and by analogy to \cite{ADFK14}). This extends the
guarantees above to imply that a transaction never reads from an aborted
transaction, and cascading aborts are avoided.

\subsubsection{Liveness}

There are two types of occurrence where an operation can wait. The first is
waiting on an access condition, or the similar condition when a transaction
attempts to commit or abort. In this case, the condition is satisfied in the
order enforced by transactions' private versions for specific objects. Since
private versions are consecutive integers and since they are acquired
atomically by the transaction, it is impossible for a circular wait to occur.
The other case of waiting is during transaction start, when private versions
are acquired. In order for this to be done atomically, transactions lock a
series of locks before getting private versions, and release the locks
afterwards. These locks are always acquired in accordance to an arbitrary
global order, regardless of transaction. That eliminates the possibility that a
circular wait occurs during start.
Since circular wait cannot occur among transactions, OptSVA-CF is
deadlock free.

\subsubsection{Progress}

Any transaction in OptSVA-CF can either abort manually or forcibly. In order
for a transaction $\tr_i$ to abort forcibly, there must be some transaction
$\tr_j$ that forces $\tr_i$ to abort, i.e., such $\tr_j$ that accessed some
object $\objx$ and released it before $\tr_i$ accessed $\objx$, and $\tr_j$
must have aborted after $\tr_i$ accessed $\objx$. Thus for every forcibly
aborted transaction, there must be another aborted transaction.
Hence, given any set of conflicting transactions, there will be at least one
transaction that will not be forcibly aborted (but it will be manually
aborted). Therefore, OptSVA-CF is strongly progressive \cite{GK10}.

\section{Architecture}
\label{sec:architecture}

\atomicrmiii{} is a framework that supplies transactional concurrency control in
a distributed system on top of Java RMI. The framework uses OptSVA-CF for synchronization.

\begin{figure}[t]
    \newcommand{\drawdashsq}[5]{%
        \fill[white]  (#1, #2) rectangle (#1+#3, #2+#4);
        \draw[loosely dashed] (#1, #2) rectangle (#1+#3, #2+#4);
        \node[above=3pt, anchor=base west] at (#1, #2+#4) {#5};
    }
    \newcommand{\drawsolidsq}[5]{%
        \fill[white] (#1, #2) rectangle (#1+#3, #2+#4);
        \draw[solid] (#1, #2) rectangle (#1+#3, #2+#4);
        \node[above=3pt, anchor=base west] at (#1, #2+#4) {#5};
    }
    \newcommand{\drawClient}[6]{
        \drawdashsq{#1+0}{#2+0}{8}{5}{{\tt Node #3}}
        \drawsolidsq{#1+0.25}{#2+0.25}{3.5}{4}{{\tt Transaction #4}}
        \drawsolidsq{#1+0.75}{#2+0.75}{3.5}{2.5}{{\tt Transactional}}    
        \drawsolidsq{#1+5.5}{#2+2.5}{2.25}{1}{{\tt Stub #5}}
        \drawsolidsq{#1+5.5}{#2+0.5}{2.25}{1}{{\tt Stub #6}}
        \draw[-latex] (#1+0.75+3.5,#2+2.5) -- ++ (.6,0) |- (#1+5.5,#2+3);
        \draw[-latex] (#1+0.75+3.5,#2+1.5) -- ++ (.6,0) |- (#1+5.5,#2+1);
        \node[] (#3exitT) at (#1+7.75,#2+3) {};
        \node[] (#3exitB) at (#1+7.75,#2+1) {};
    }
    \newcommand{\drawServer}[6]{
        \drawdashsq{#1+0}{#2+0}{7.5}{5.5}{{\tt Node #3}}
        \drawsolidsq{#1+0.25}{#2+4.5}{5}{.5}{{\tt RMI Registry}}
        \drawsolidsq{#1+3.75}{#2+0.75}{3.5}{2.5}{{\tt Shared Object #6}}    
        \drawsolidsq{#1+0.25}{#2+2.5}{2.25}{1}{{\tt Proxy #4}}
        \drawsolidsq{#1+0.25}{#2+0.5}{2.25}{1}{{\tt Proxy #5}}
        \draw[-latex] (#1+0.25+2.25,#2+3) -- ++ (.6,0) |- (#1+3.75, #2+2.5);
        \draw[-latex] (#1+0.25+2.25,#2+1) -- ++ (.6,0) |- (#1+3.75,#2+1.5);
        \node[] (#3inputT) at (#1+0.25,#2+3) {};
        \node[] (#3inputB) at (#1+0.25,#2+1) {};
    }
    \scalebox{.5}{
    \begin{tikzpicture}
        \drawClient{-9.5}{6}{C1}{T1}{T1/A}{T1/B}
        \drawClient{-9.5}{0}{C2}{T2}{T2/A}{T2/B}
        \drawServer{0}{6}{S1}{T1/A}{T2/A}{A}
        \drawServer{0}{0}{S2}{T1/B}{T2/B}{B}
    
        \draw[-latex] (C1exitT.center) -- (S1inputT.center);
        \draw[-latex] (C1exitB.center) -- ++ (.6,0) -- ++ (.8,-4) -- (S2inputT.center);
        \draw[-latex] (C2exitT.center) -- ++ (.6,0) -- ++ (.8,4) -- (S1inputB.center);
        \draw[-latex] (C2exitB.center) -- (S2inputB.center);
    \end{tikzpicture}
    }
\caption{\label{fig:architecture} \atomicrmiii{} architecture (2 transactions, 2 objects).}
\end{figure}
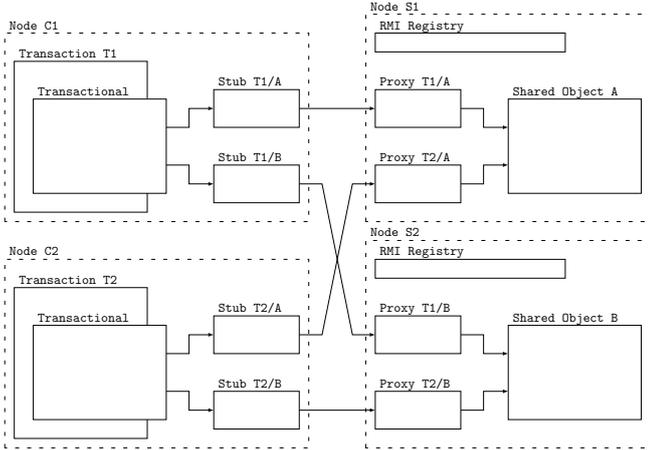

We give an overview of the system architecture in \rfig{fig:architecture} with
two client nodes and two server nodes, showing the flow of control when remote
methods are executed on shared objects.
In general, the system may contain any number of independent client and server
nodes. Each of the server nodes hosts a number of discrete, uniquely
identifiable shared objects, whose methods are called by client nodes as
operations. Any node can act simultaneously as client as well as a server.
Each shared object is located at exactly one specific node (as opposed to the
object being copied or moved to other nodes, or being replicated on several
nodes) and all operations invoked on that object will execute on the node that
hosts it. 

\begin{figure}
\begin{lstlisting}[numbers=none]
interface Account extends Remote {
  @Access(Mode.READ)   int  balance();
  @Access(Mode.UPDATE) void deposit(int value);
  @Access(Mode.UPDATE) void withdraw(int value);
  @Access(Mode.WRITE)  void reset();
}
\end{lstlisting}
    \caption{\label{fig:interface}Shared object interface example (a bank
    account).} 
\end{figure}

\begin{figure}
    \begin{lstlisting}[numbers=none]
interface Transaction {
  Transaction(boolean irrevocable);

  <T> T updates(T obj);
  <T> T writes(T obj);
  <T> T reads(T obj);
  <T> T accesses(T obj);
  
  <T> T updates(T obj, int maxUpdates);
  <T> T writes(T obj, int maxWrites);
  <T> T reads(Tobj, int maxReads);
  <T> T accesses(T obj, int maxRd, int maxWr, int maxUpd);

  void start(Transactional runnable);
  void commit();
  void retry();
  void abort();
}

interface Transactional {
  void atomic(Transaction t);
}
\end{lstlisting}
\caption{\label{fig:transaction-interface}\atomicrmiii{} transaction interface.} 
\end{figure}

Shared objects can be accessed remotely from client nodes in the system by
calling methods specified by the object's interface (as per the control flow
model). 
A shared objects interface is defined by the programmer and consists of a
number of methods that can be called on the object. The semantics of the
operations are defined by the programmer and can be anything from simple gets
and sets, to complex methods executing arbitrary server-side code, accessing a
database, or even invoking other remote objects.
Each method must be annotated as either a read operation, a write or an
update operation. 
We give an example of a shared object interface for a bank account in
\rfig{fig:interface}.

\begin{figure}
    \begin{lstlisting}[numbers=none]
  Transaction t = new Transaction(irrevocable=false);

  Account a = t.accesses(registry.locate("A"), 1, 0, 1);
  Account b = t.updates(registry.locate("B"), 1);
    
  t.start(new Transactional() {
    void atomic(Transaction t) {
      a.withdraw(100);
      b.deposit(100);
      if (a.balance() < 0)
        t.abort();
    }
  });    
  \end{lstlisting}
    \caption{\label{fig:transaction-abort}Transaction definition example (with manual abort).} 
\end{figure}

Clients execute operations on shared objects as part of transactions. We show
an example of a transaction definition in \rfig{fig:transaction-abort}.
The programmer declares a transaction using the API provided by \atomicrmiii{}
(\rfig{fig:transaction-interface}), by creating a {\tt Transaction} object,
which is responsible for starting and stopping transactional execution. 
A transaction can be defined as irrevocable at this point, meaning it will
never be forced to abort, because it will not access objects that are released
early, instead waiting for the preceding transaction to commit or abort.
Then,
that object is used to declare the transaction's preamble, where the programmer
specifies which objects will be used by the transaction and how, by passing the
reference retrieved from the RMI registry to method {\tt reads}, {\tt updates},
{\tt writes}, or {\tt accesses}---the latter if more than one kind of operation
may be executed on the object. 
The programmer can use variants these methods to also provide \emph{suprema}
for any object used by the transaction. The suprema indicate the maximum number
of times the transaction will execute read-only, write, and update methods on each
shared object throughout the execution of the code. 
In the example in \rfig{fig:transaction-abort} the preamble declares the
transaction will invoke at most one read-only, at most one update method on
{\tt A}, and at most one update method on  shared object {\tt B}.

In practice, suprema do not have to be derived manually, but instead static
analysis \cite{SW12} or the type system \cite{Woj05b} can be used. If suprema are not
given, infinity is assumed (and the system maintains guarantees). If suprema
are provided though, the underlying concurrency control algorithm uses them to
effect early release, and in this way increase the level of parallelism between
concurrent transactions.

\subsection{Instrumentation}

When accesses are declared within the preamble, an object stub is created. This
stub is then used within the code of the transaction to invoke methods on the
shared objects, as with ordinary RMI stubs. The difference between an ordinary
RMI stub and an \atomicrmiii{} stub is that the latter does not forward method
calls to the shared object directly, but instead uses a proxy object.  Proxy
objects are created dynamically on the node hosting the shared object in
question at the same time as the stub is created by the transaction.  Each
proxy object links one specific shared object on the server side with one
specific transaction (object) on the client side. Proxy objects implement the
interfaces of the shared objects they are linking, and their role is to inject
the concurrency control code of OptSVA-CF before and after the invocation of
specific methods of the shared object. The injection is done via reflection,
which supplies the necessary flexibility, allowing arbitrary methods to easily
be supplemented with concurrency control.
In theory, proxy objects could be located either on the server side or the
client side, but since the communication between the proxy and the shared
object is much more frequent than that between the transaction and the proxy,
placing them on the server-side incurs lower overheads.
In addition, if the proxy is placed on the server, it can easily manage copy
and log buffers, which must be placed on the server to preserve the CF
model---methods executed using buffers should have side effects on the same
node as the original object.
Proxy objects can be decommissioned once a transaction that created them
finishes executing.

\subsection{Transactional Code}

Once the preamble of the transaction is in place, the transaction's code can be
specified.  This is done by creating an object implementing the {\tt
Transactional} interface, whose {\tt atomic} method then defines the logic of
the transaction.
In general, a transaction can contain virtually any code between its start and
either commit or abort. This specifically means that apart from operations on
shared objects, any local operations, e.g., irrevocable operations, can be
present within.
As an example of a simple transaction, in \rfig{fig:transaction-abort} the
programmer specifies a transaction that transfers 100 currency from one bank
account to another. Thus, an anonymous {\tt Transactional} object is created
and within the {\tt atomic} method, the {\tt withdraw} method is called on
object {\tt a} (the stub for shared object {\tt A}), after which the {\tt
deposit} method is called on {\tt b} (the stub for {\tt B}). The programmer can
rest assured the concurrency control algorithm will synchronize the execution
of this code so that no other transaction in the system interferes with the
execution in a way that would violate its consistency.  If the transaction
reaches the end of its code it attempts to commit.  The programmer is also
given the option to abort or retry the entire transaction manually by using the
transaction object and invoking either the {\tt abort} or {\tt retry} method.
Here, the transaction is rolled back if it turns out that the balance on
account {\tt A} fell below 0 as a result of executing {\tt withdraw}. 
Since \atomicrmiii{} implements a pessimistic concurrency
control algorithm, transactions never abort 
as a result of
conflict (see discussion in \rsec{sec:optsva-cf}).
It is possible for a transaction to abort as a result of explicitly invoking
{\tt abort}, which can cause a cascading rollback, or if a failure is detected
(discussed below). 
Any transaction can be prevented from ever aborting though by specifying it as irrevocable.

\subsection{Executor Thread}

OptSVA-CF calls for asynchronous execution using separate threads. Given the
cost of overhead that starting a thread creates, \atomicrmiii{} uses one
executor thread per JVM. The executor thread is always running and transactions
assign it tasks. Each task consists of a condition and code. The code of the
task is meant to be executed only when the condition is satisfied. Once the
thread receives a task, it checks whether it can be immediately executed. If
not, it queues up the task and waits until any of the two counters that can
impact the condition change value ($\lvf$ and $\ltvf$). When any of the
counters change, the thread re-evaluates the relevant conditions and executes
the task, if the condition so allows.

\subsection{Fault Tolerance Mechanisms}

In distributed environments faults are a fact of life, so any DTM system must
have mechanisms to deal with them. \atomicrmiii{} handles two basic types of
failures: remote object failures and transaction failures.

Remote object failures are straightforward and the responsibility for
detecting them and alarming \atomicrmiii{} falls onto the mechanisms built into
Java RMI. Whenever a remote object is called from a transaction and it cannot
be reached, it is assumed that this object has suffered a failure and an
exception is thrown. The programmer may then choose to handle the exception
by, for example, rerunning the transaction, or compensating for the failure.
Remote object failures follow a \emph{crash-stop} model of failure: any object 
that crashed is removed from the system.

On the other hand, a client performing some transaction can crash causing a
transaction failure. Such failures can occur before a transaction releases all
its objects and thus make them inaccessible to all other transactions. The
objects can also end up in an inconsistent state. For these reasons transaction
failures need also to be detected and mitigated. \atomicrmiii{} does this by having
remote objects check whether a transaction is responding. If a transaction
fails to respond to a particular remote object (times out), it is considered to
have crashed, and the object performs a rollback on itself: it reverts its
state and releases itself. If the transaction actually crashed, all of its
objects will eventually do this and the state will become consistent. On the
other hand, if the crash was illusory and the transaction tries to resume
operation after some of its objects rolled themselves back, the transaction
will be forced to abort when it communicates with one of these objects.

\section{Evaluation}
\label{sec:evaluation}

In this section we present the results of a practical evaluation of \atomicrmiii{}
in the context of other distributed TM
concurrency control mechanisms operating in a similar system model. 

\subsection{Frameworks}

The first framework we use for evaluation is Atomic RMI \cite{SW15-ijpp}, a
distributed pessimistic CF TM implementing SVA \cite{Woj05b} on top of Java
RMI. SVA uses the bare supremum versioning mechanism described in
\rsec{sec:supremum-versioning} and is operation-type agnostic. The comparison
against SVA shows off the optimizations introduced in OptSVA-CF, especially
since the two algorithms are implemented using the same technology and give the
same guarantees.
The second framework we compare \atomicrmiii{} against is HyFlow2 \cite{TRP13}, a
state-of-the-art distributed TM system implemented in Scala.
HyFlow2 implements the optimistic Transactional Forwarding Algorithm (TFA)
\cite{SR11} and operates in the data flow model.
TFA is opaque but does not have provision for irrevocable operations.

We also compare all three TM systems against distributed concurrency control
solutions based on locks. Specifically, we use distributed mutual exclusion
locks (marked Mutex) and read-write locks (R/W Locks), both custom-tailored and
implemented on top of Java RMI. In both solutions a lock is created for
every shared object in the system. 
Each locking solution has two variants.
The first variant is a straightforward usage where every transaction locks
every object from its access set when it commences, and releases each of object
on commit. This is equivalent to a \emph{conservative (strong) strict two-phase
locking} solution and satisfies opacity.
We denote this variant as \emph{S2PL}.
The second variant represents \emph{non-strict two-phase locking} (\emph{2PL}),
and is a more advanced implementation from the programmer's point of view.
Here, each transaction also initially locks each of the objects in its access
set, but the programmer determines the last access on each object in the
transaction's access set and manually releases the lock early (prior to
commit). 
Non-strict two-phase locking satisfies last-use opacity under the assumption
that last access is always determined correctly.
We denote the second variant as \emph{2PL}.
Finally, we also use a solution with a \emph{single global mutual exclusion lock}
(\emph{GLock}) that is acquired by each transaction for the duration of the
transaction's entire execution. This produces a completely sequential execution
and acts as a baseline for the purpose of the comparison.

\subsection{Benchmark}

We perform our evaluation using a 16-node cluster connected by a 1Gb network.
Each node had two quad-core Intel Xeon L3260 processors at 2.83
GHz with 4 GB of RAM each and runs OpenSUSE 13.1 (kernel 3.11.10, x86\_64
architecture). We use Groovy version 2.3.8  with the 64-bit Java HotSpot(TM)
JVM version 1.8 (build 1.8.0\_25-b17).

The evaluation is performed using our own distributed implementation of
Eigenbench \cite{SOJB+10}. 
Eigenbench is a flexible, powerful, and lightweight benchmark that can be used
for comprehensive evaluation of mutlicore TM systems by simulating a variety of
transactional application characteristics.

Eigenbench uses three arrays of shared objects, each of which is accessed
with a different level of contention. The hot array contains some number $n$ of
objects that can be accesses by transaction in any thread. The access to
objects in the hot array is controlled by the TM. The mild array contains $n$
objects per thread. The access to these objects is also controlled by the TM,
but the objects are partitioned in such a way, that no two transactions ever
conflict on them. The third, cold array is populated and partitioned like the
mild array, but it is only accessed non-transactionally.
Each object within any of the three arrays is a reference cell, i.e., an object
that holds a single value, that can be either read or written to.
These arrays are accessed by client transactions.  Each transaction accesses
semi-randomly selected objects in all three arrays in random order, with the
exception that the number of accesses to each type of array is specified, and
the ratio of read operations to write operations on each type of array is also
specified. 
The benchmark has a specified locality, which is a probability with which
transactions will access the same object several times. When an object is being
selected by a transaction, a random number is generated, and if it is below the
locality probability, the object is selected at random from the transaction's
history of objects accessed thus far. Otherwise, the object is selected
randomly from the pool of all shared objects. Locality and the length of the
history are parameters of the benchmark.

\subsection{Results and Discussion}

\begin{figure*}

\begin{subfigure}[b]{.33\linewidth}
    {\includegraphics[width=\linewidth]{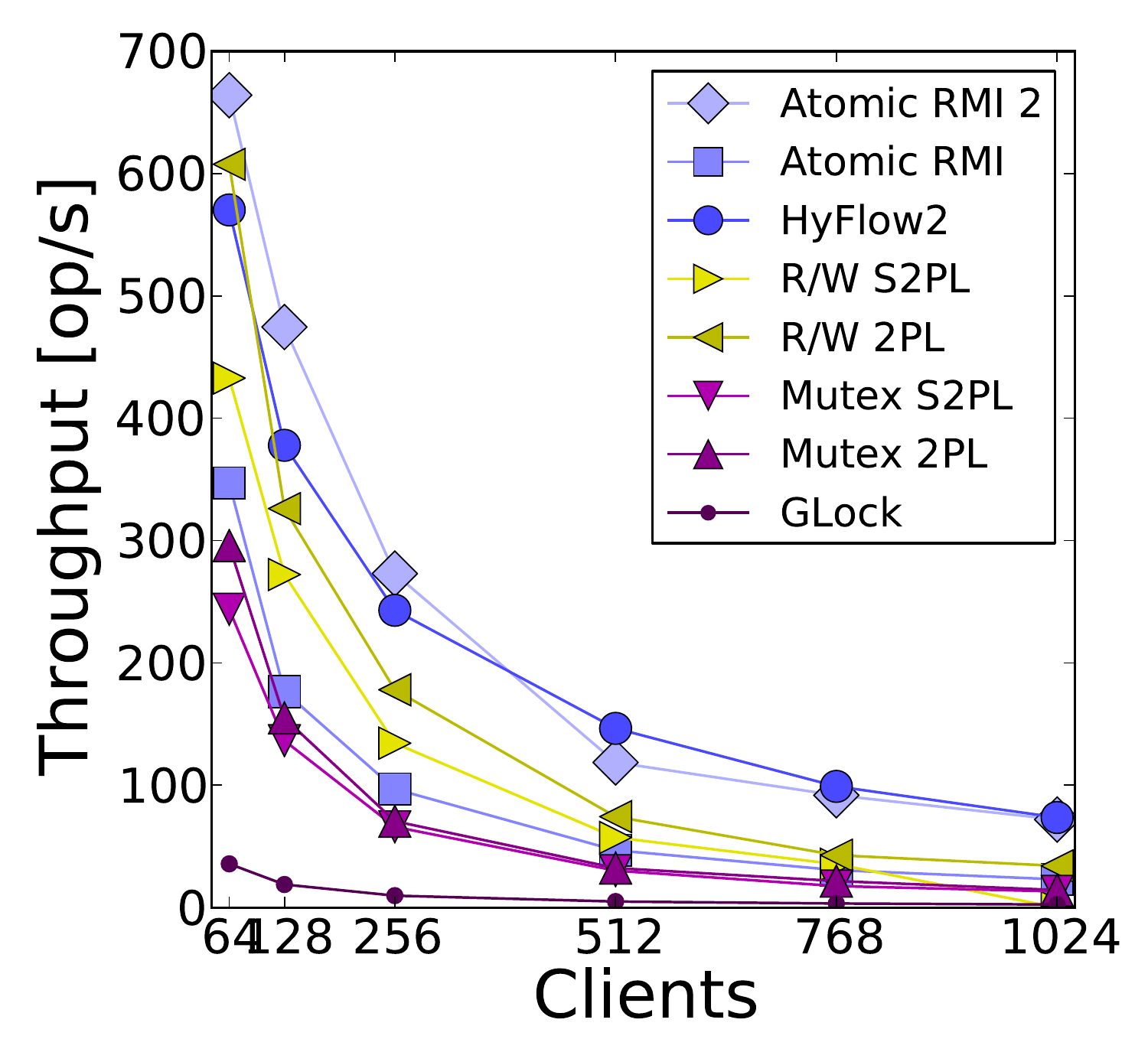}}
    \caption{\label{fig:scalability-read}90\% reads, 10\% writes.}
\end{subfigure}
\begin{subfigure}[b]{.33\linewidth}
    {\includegraphics[width=\linewidth]{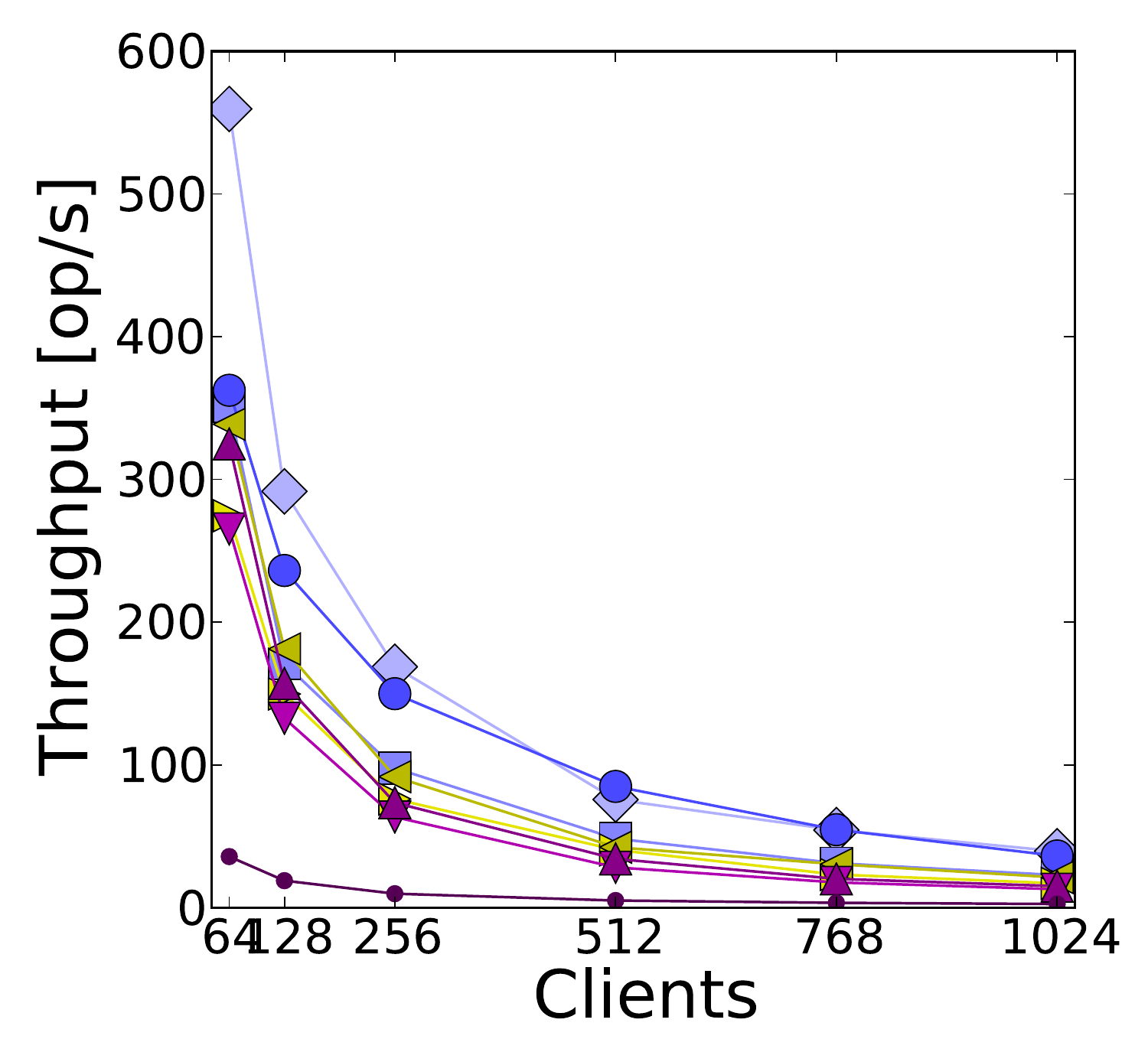}}
    \caption{\label{fig:scalability-balanced}50\% reads, 50\% writes.}
\end{subfigure}
\begin{subfigure}[b]{.33\linewidth}
    {\includegraphics[width=\linewidth]{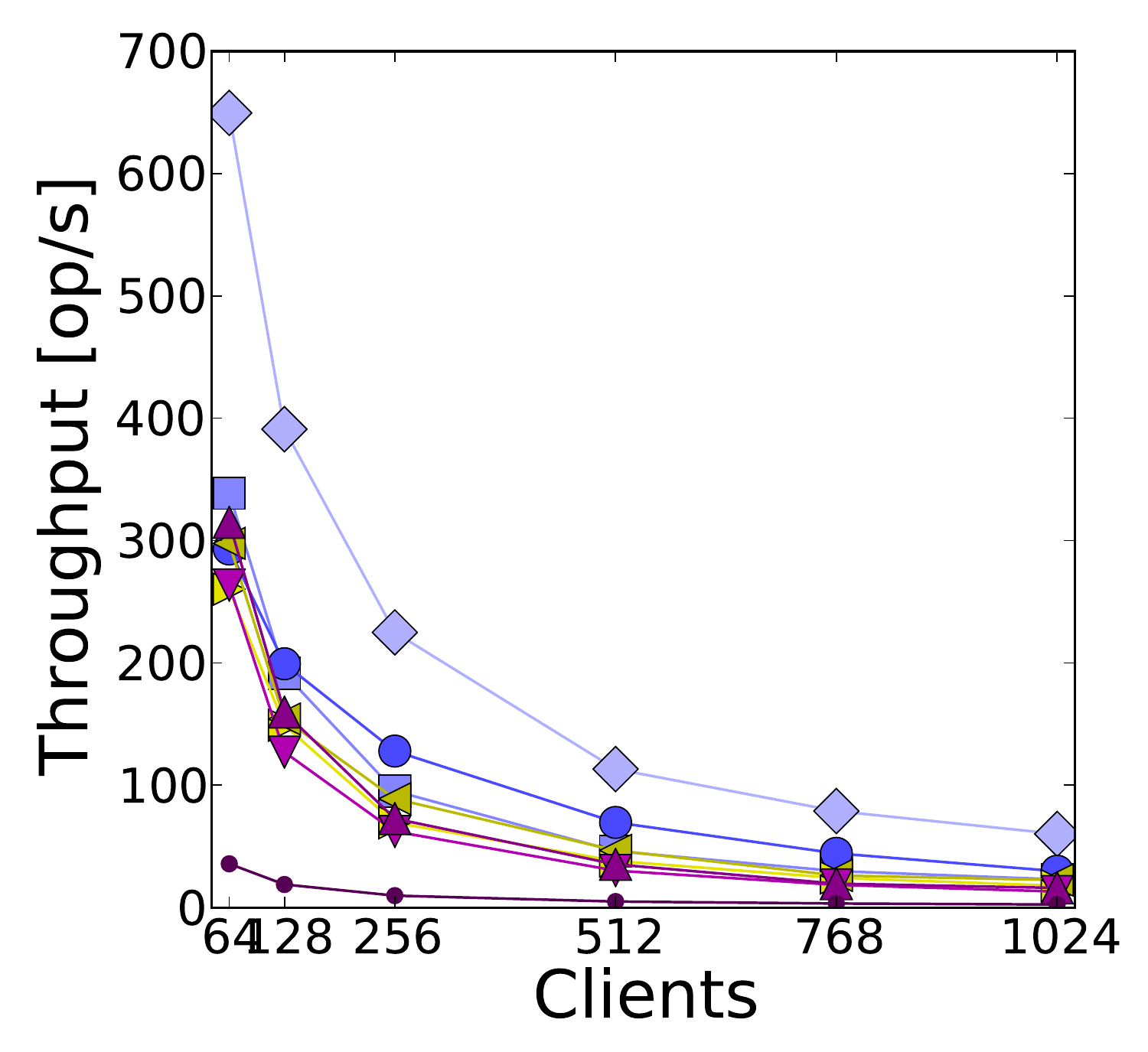}}
    \caption{\label{fig:scalability-write}10\% reads, 90\% writes.}
\end{subfigure}
\caption{\label{fig:scalability}Throughput vs client count.}
\end{figure*}
\begin{figure*}
\begin{subfigure}[b]{.33\linewidth}
    {\includegraphics[width=\linewidth]{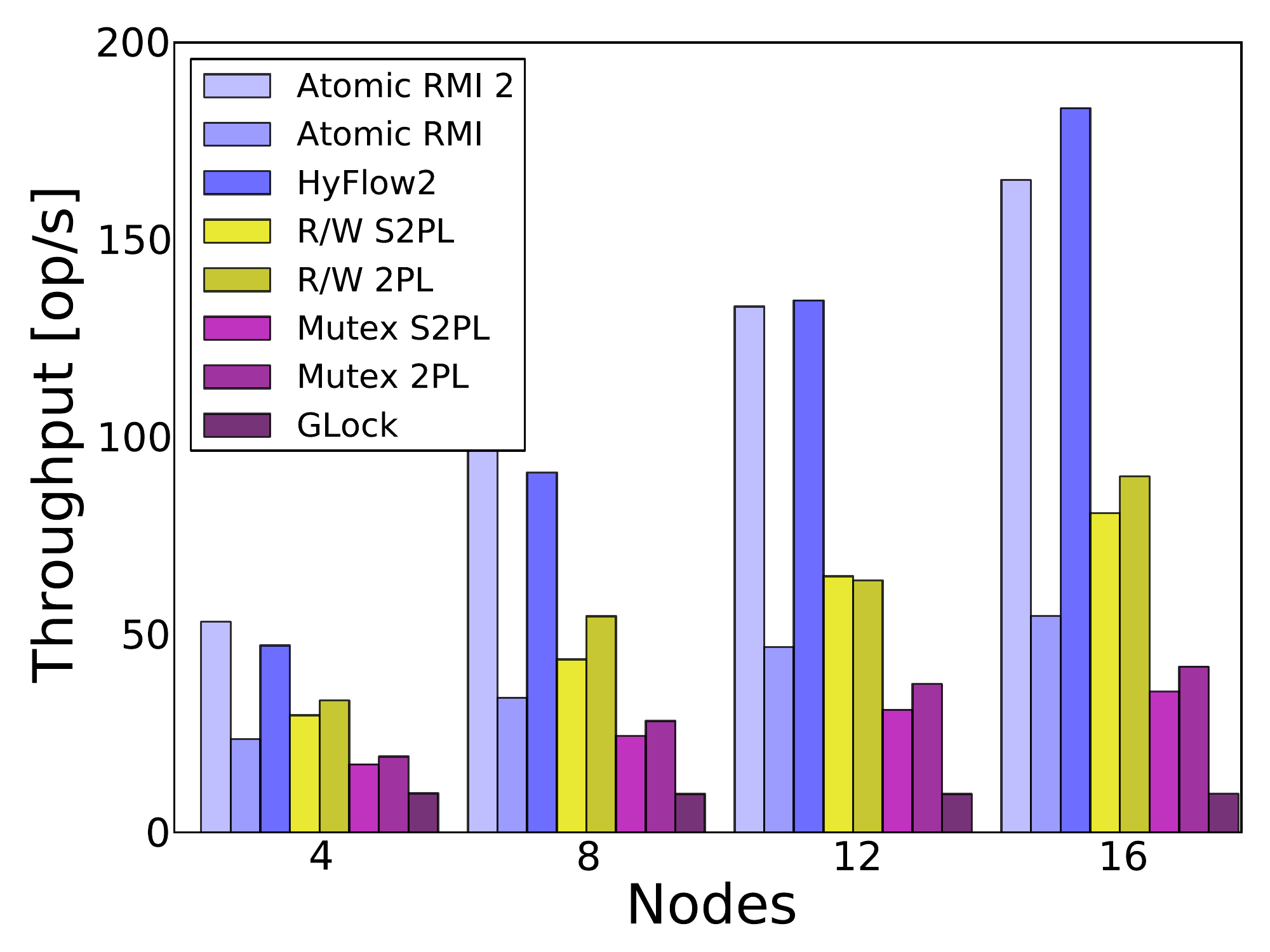}}
    \caption{\label{fig:throughput-hot-read-5}90\% reads, 10\% writes, 5 arrays.}
\end{subfigure}
\begin{subfigure}[b]{.33\linewidth}
    {\includegraphics[width=\linewidth]{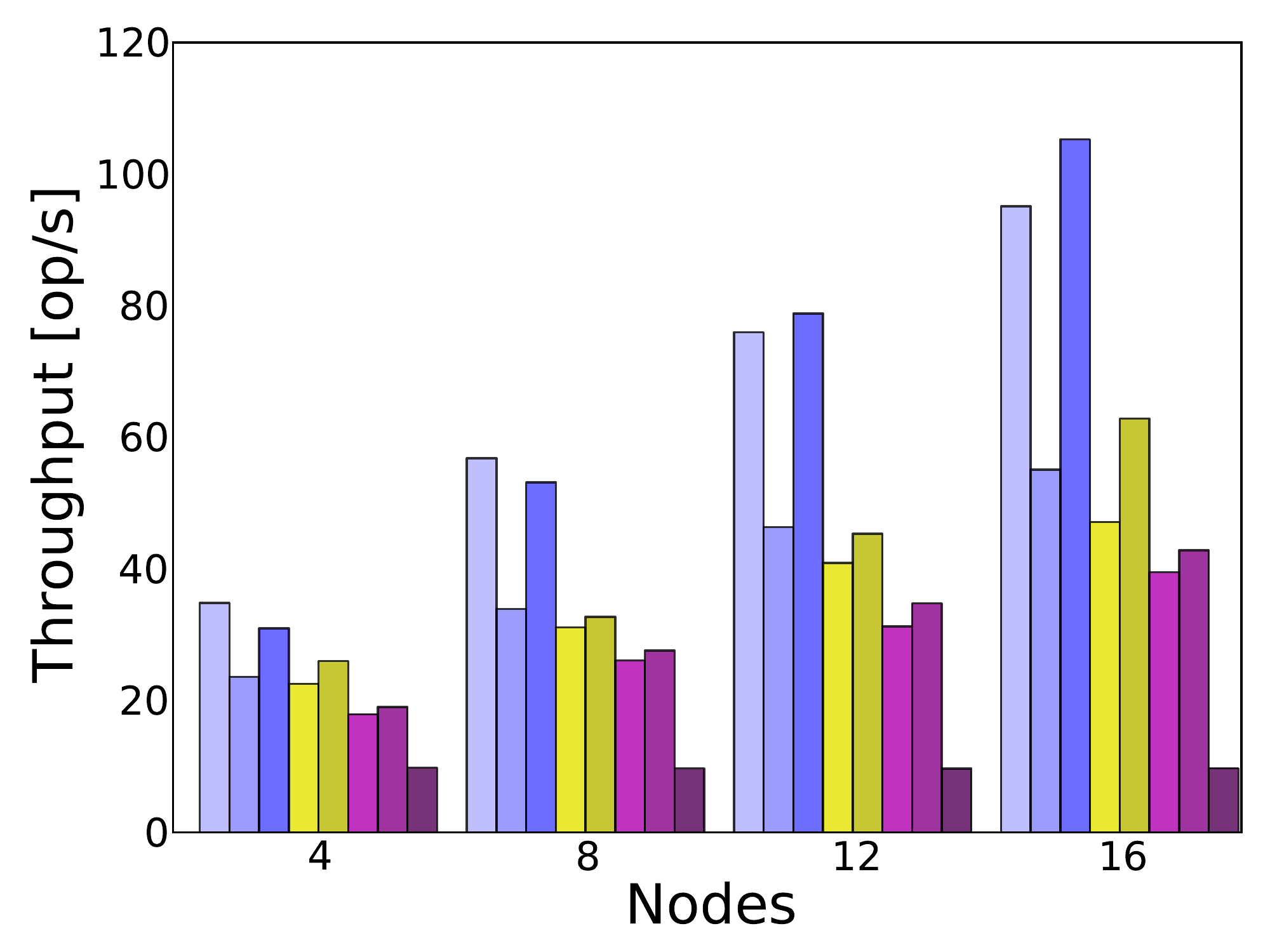}}
    \caption{\label{fig:throughput-hot-balanced-5}50\% reads, 50\% writes, 5 arrays.}
\end{subfigure}
\begin{subfigure}[b]{.33\linewidth}
    {\includegraphics[width=\linewidth]{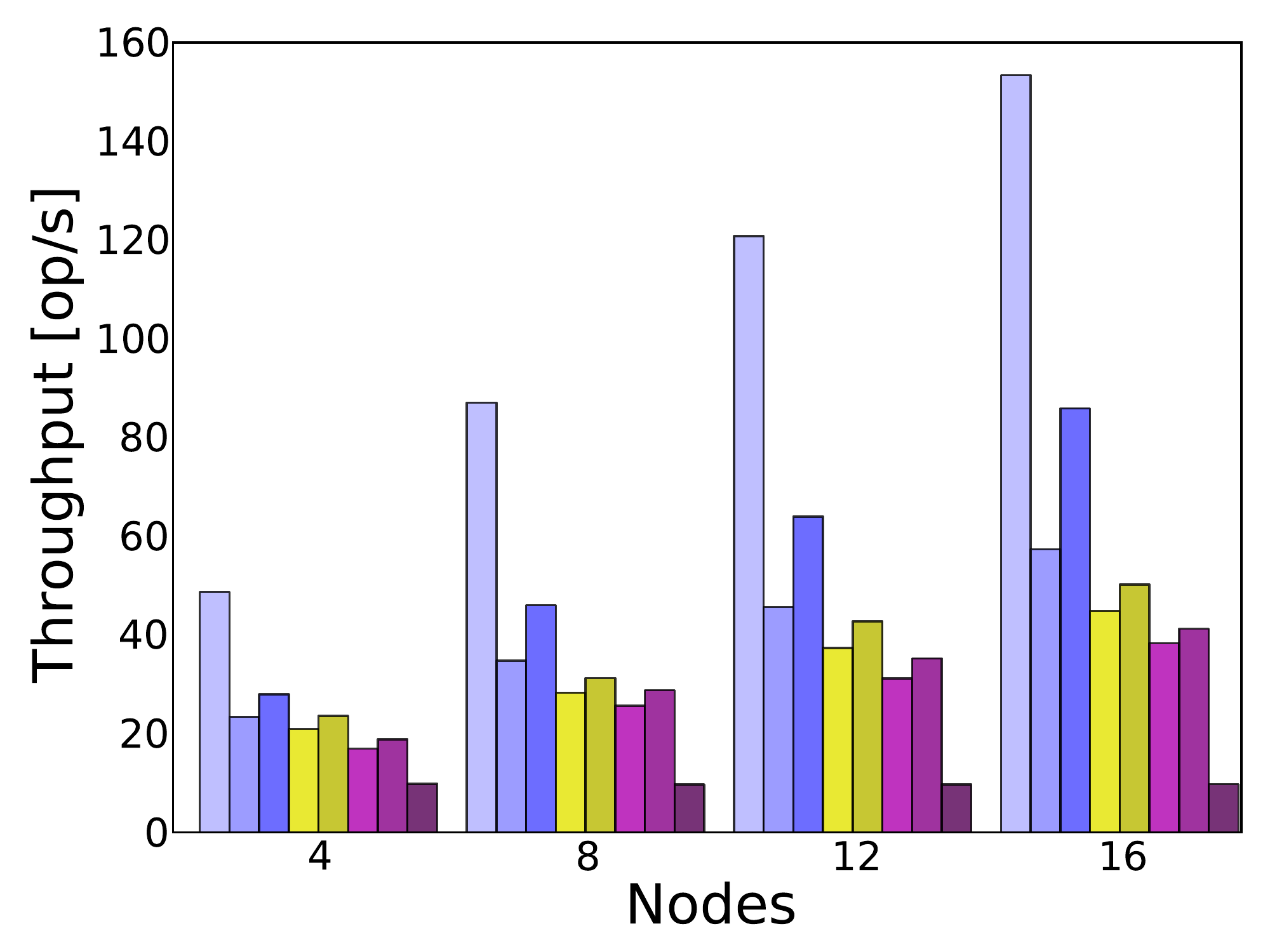}}
    \caption{\label{fig:throughput-hot-write-5}10\% reads, 90\% writes, 5 arrays.}
\end{subfigure}
\begin{subfigure}[b]{.33\linewidth}
    {\includegraphics[width=\linewidth]{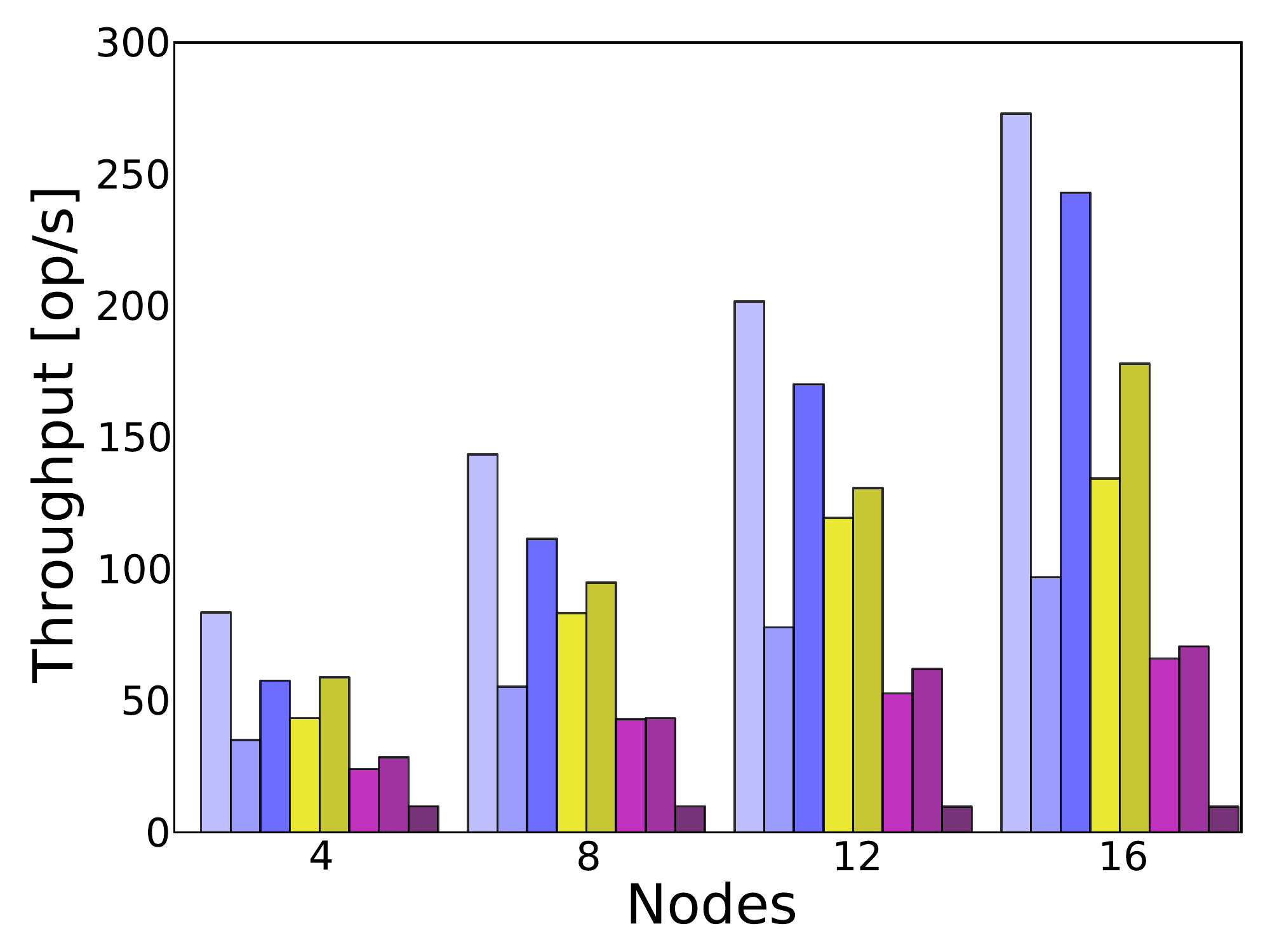}}
    \caption{\label{fig:throughput-hot-read-10}90\% reads, 10\% writes, 10 arrays.}
\end{subfigure}
\begin{subfigure}[b]{.33\linewidth}
    {\includegraphics[width=\linewidth]{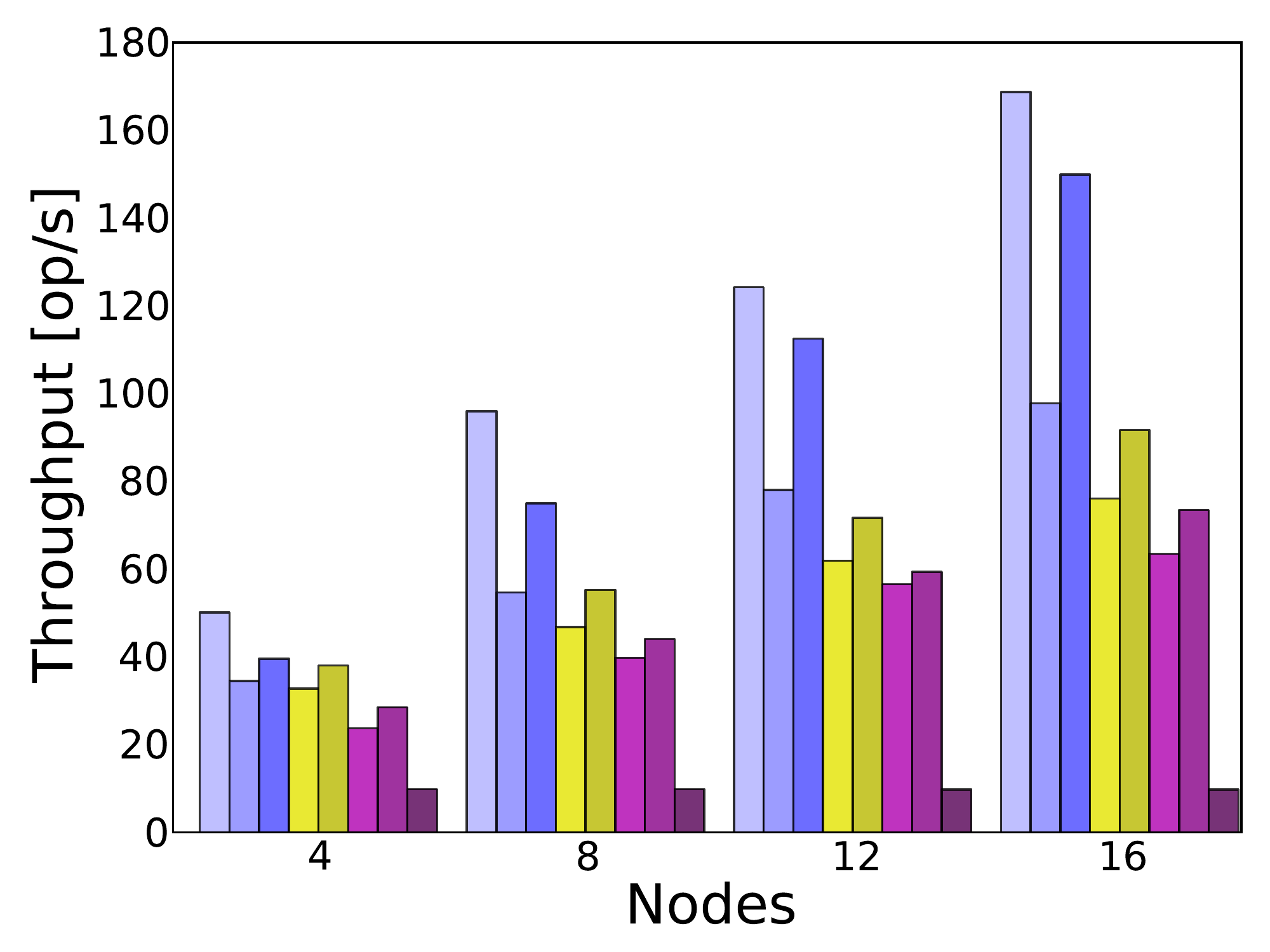}}
    \caption{\label{fig:throughput-hot-balanced-10}50\% reads, 50\% writes, 10 arrays.}
\end{subfigure}
\begin{subfigure}[b]{.33\linewidth}
    {\includegraphics[width=\linewidth]{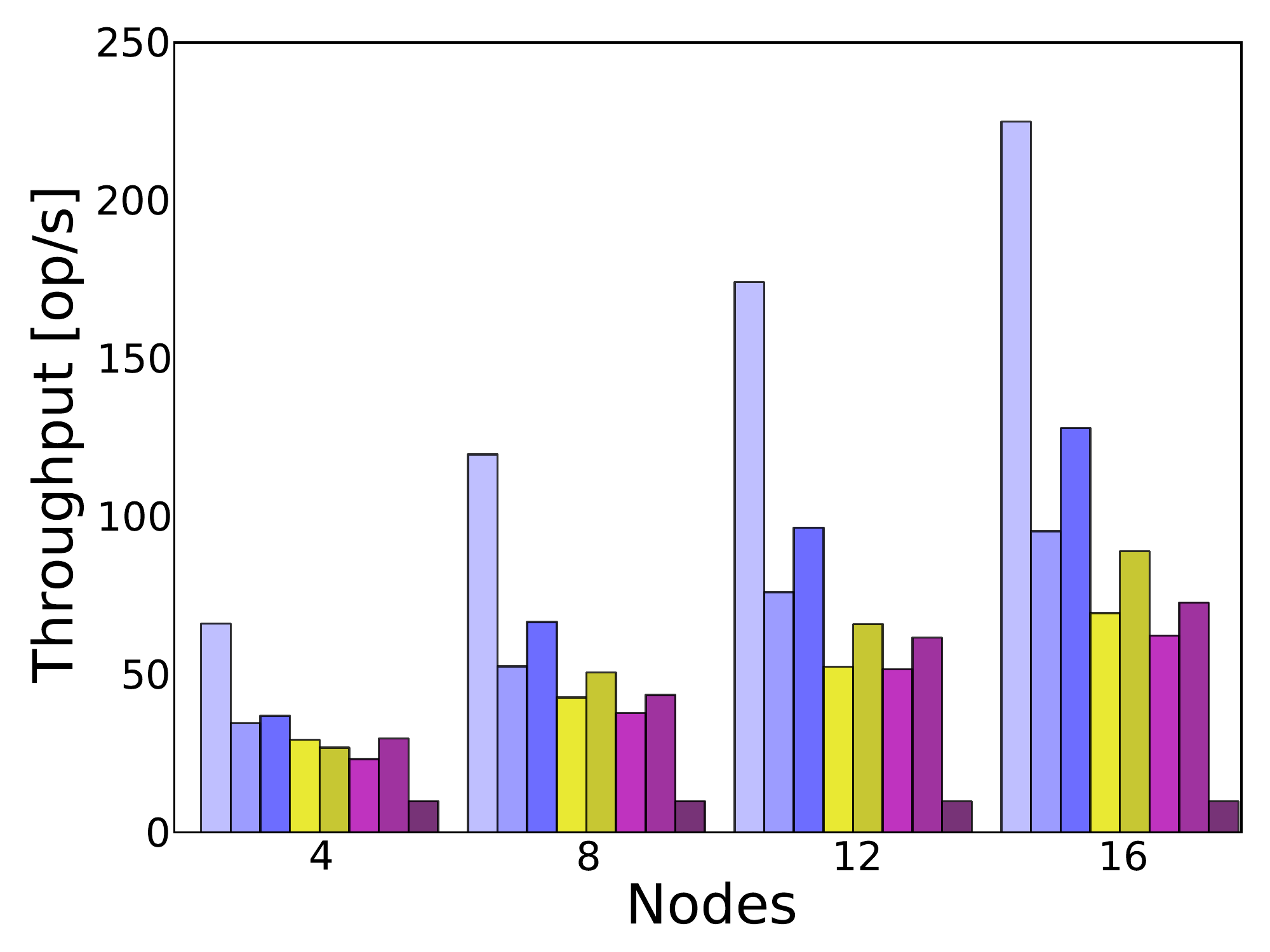}}
    \caption{\label{fig:throughput-hot-write-10}10\% reads, 90\% writes, 10 arrays.}
\end{subfigure}
\caption{\label{fig:throughput-hot}Throughput vs node count (hot array accesses).}
\end{figure*}
\begin{figure*}
\begin{subfigure}[b]{.33\linewidth}
    {\includegraphics[width=\linewidth]{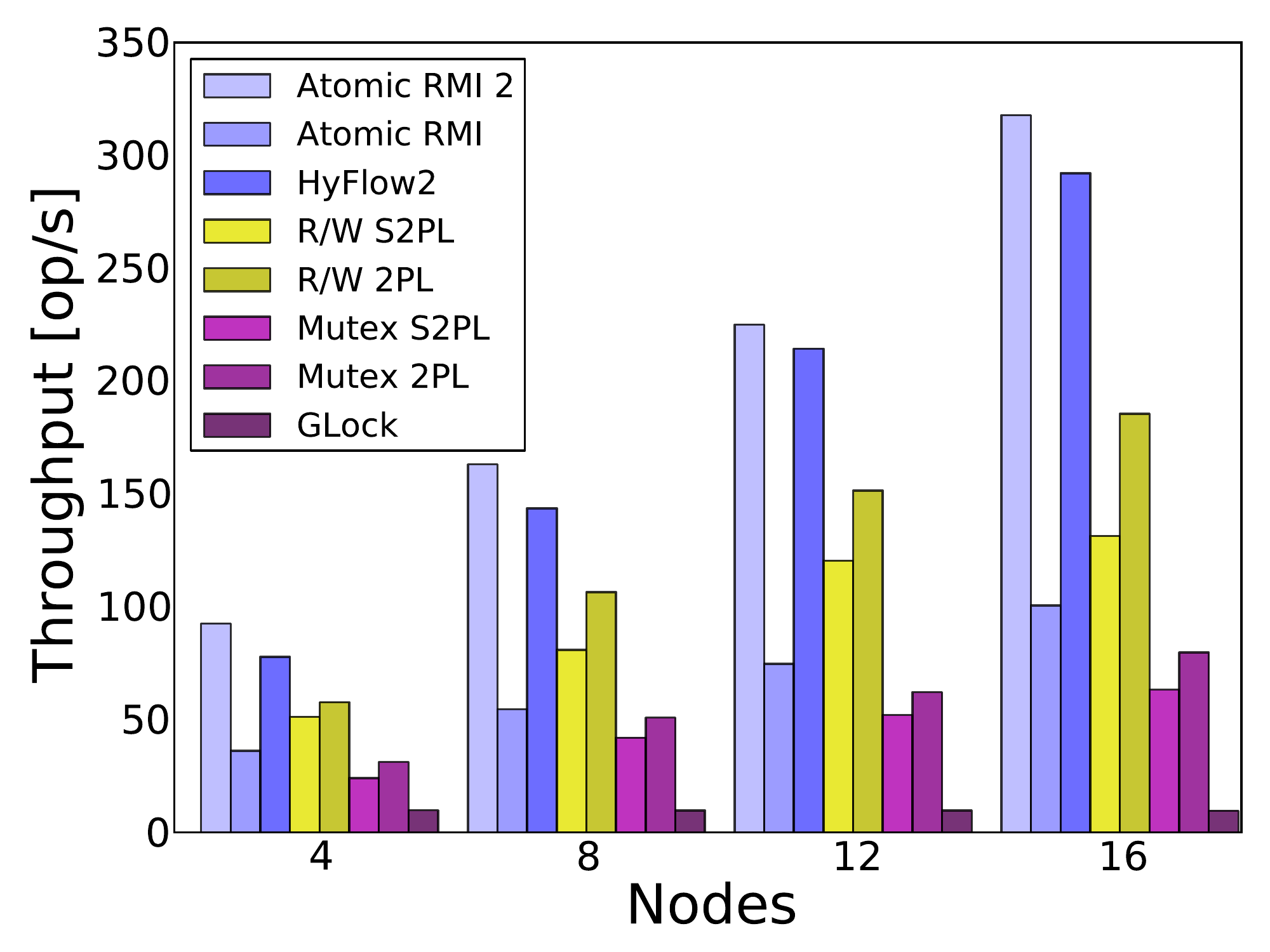}}
    \caption{\label{fig:throughput-hotmild-read}90\% reads, 10\% writes.}
\end{subfigure}
\begin{subfigure}[b]{.33\linewidth}
    {\includegraphics[width=\linewidth]{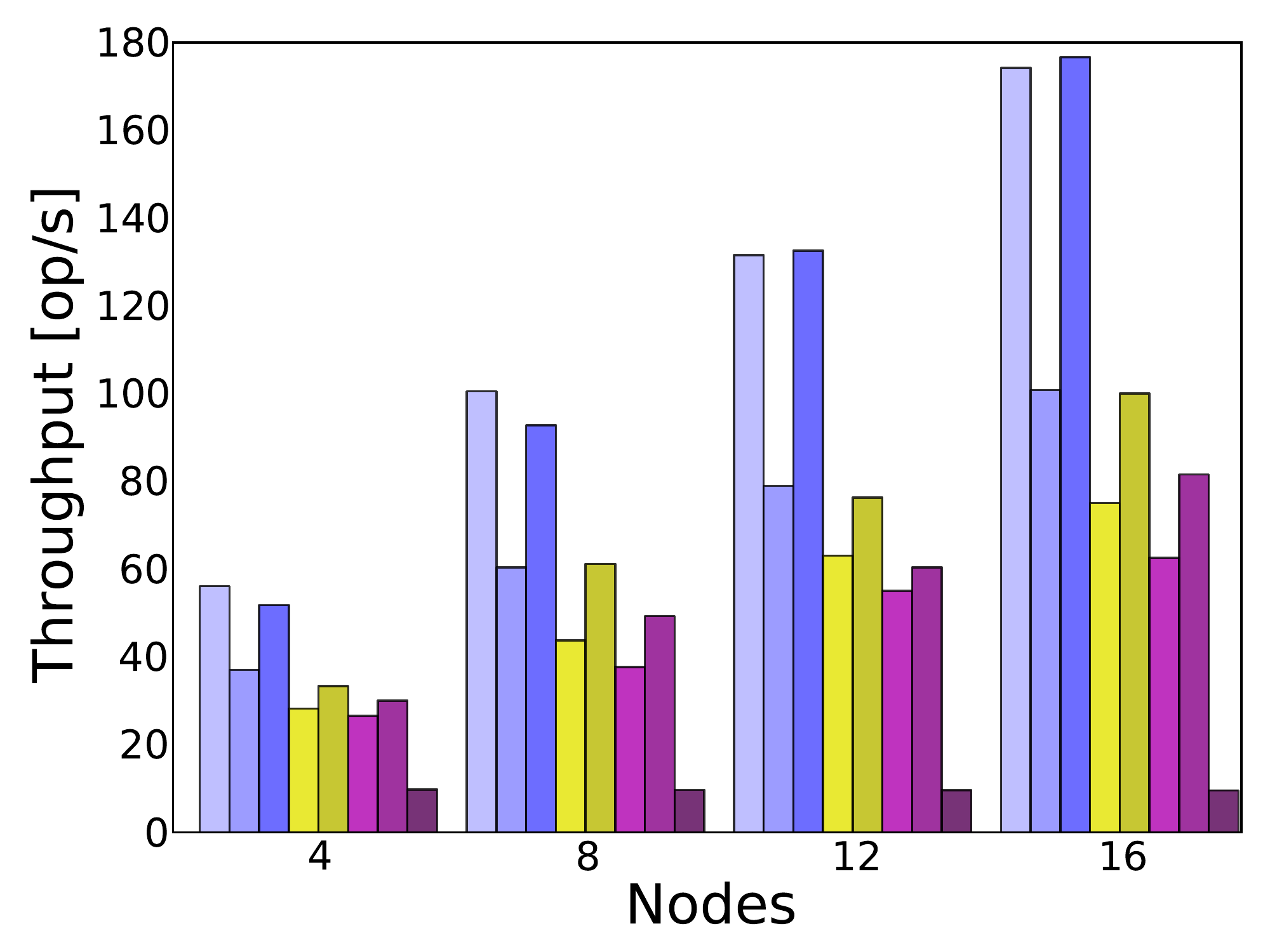}}
    \caption{\label{fig:throughput-hotmild-balanced}50\% reads, 50\% writes.}
\end{subfigure}
\begin{subfigure}[b]{.33\linewidth}
    {\includegraphics[width=\linewidth]{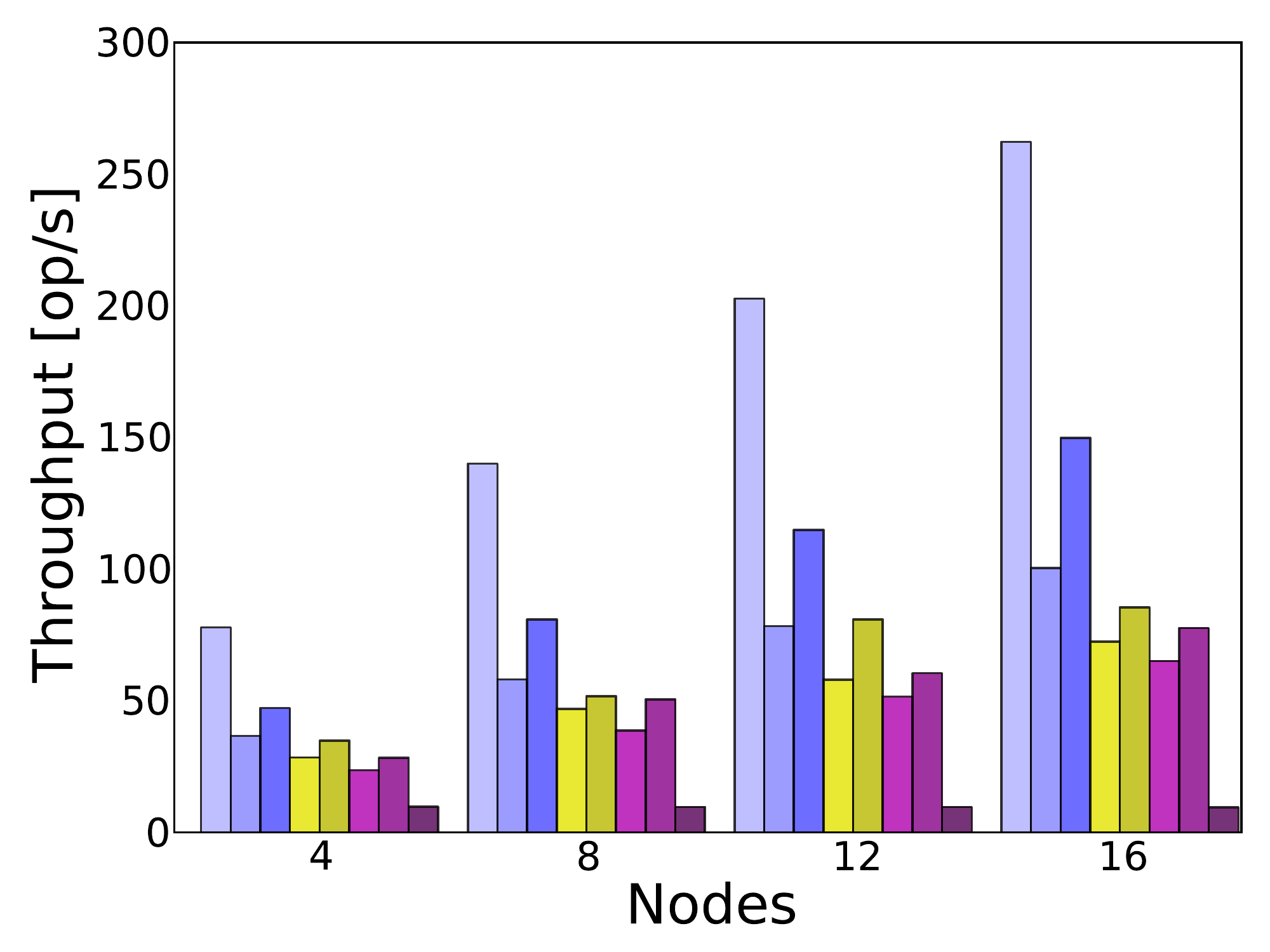}}
    \caption{\label{fig:throughput-hotmild-write}10\% reads, 90\% writes.}
\end{subfigure}
\caption{\label{fig:throughput-hotmild}Throughput vs node count (hot and mild array accesses).}
\end{figure*}

\rfig{fig:scalability} illustrates the change of throughput (measured in the
number of executed operations on shared data per second) as the number of
clients increases from 64 (4 per node) to 1024 (64 per node). We show three
scenarios, each executed on 16 nodes, with 10 arrays of each type per node.
Each client executes 10 consecutive transactions, each with 10 operations on
the hot array per transaction, with a 9$\div$1, 5$\div$5, or 1$\div$9
read-to-write operation ratio. Each operation takes around 3ms to execute, not
counting the overhead from synchronization, network communication, or
serialization overhead. This means operations are fairly long, which represents
the complex computations. The locality of operations
is set to 50\% with a history of 5 operations.

The graphs show that all frameworks' throughput falls as the number of clients,
and therefore contention, increases. The decline is steep until 256 clients,
and it levels out by 1024 clients. All systems significantly outperform the
serial execution forced through GLock.
In the 90\% read scenario HyFlow2 and \atomicrmiii{} outperform other frameworks by a
significant margin of between 9 and 267\% (not counting GLock), with
the exception of R/W 2PL outperforming HyFlow2 at 64 clients. \atomicrmiii{}
outperforms HyFlow2 initially (by 9--25\%), but after 512 clients are
introduced, HyFlow2 takes the lead (by 2--23\%), and both frameworks
throughputs eventually converge at the 1024 client mark.
In the other two scenarios, all frameworks suffer a decrease in throughput, but
\atomicrmiii{} remains relatively efficient, outperforming all other frameworks,
including HyFlow2, by 9--359\%.
The difference stems from the write-oriented optimizations in \atomicrmiii{} that allow
the framework to tighten the executions in the presence of larger contingents of
write operations, just as much as is possible in read-dominated schedules:
objects are acquired for writing as late as possible and released prior to commit.
Meanwhile other frameworks typically do not optimize write operations to the
same extent. Specifically, HyFlow2 does not release early on writes, and R/W
2PL cannot perform any optimizations on writes, apart from early release on
last write.
In addition a degradation in \atomicrmiii{}'s performance is also partly explained by
the need to introduce new threads to handle asynchrony, which can become a
bottleneck and offset the gain from \atomicrmiii{}'s optimizations if other threads are
also running on the same node (like client threads here).
Among the remaining frameworks, any 2PL always performs better than the
apposite S2PL variant, and R/W performs better than Mutex. Atomic RMI performs on par
with Mutex 2PL and significantly below \atomicrmiii{}.

\rfig{fig:throughput-hot} shows a change in throughput with constant contention
as new nodes are introduced. In this scenario, we vary the number of nodes from
4 to 16 with 5 or 10 arrays of each type hosted on each node (yielding lower
and higher contention respectively), and 16 clients running per node. The
remainder of parameters is as above.
As more processors are introduced into the system, the number of transactions
running in parallel increases, causing the throughput of all frameworks to
increase as well.

In the 5-array scenarios in
\rfig{fig:throughput-hot-read-5}--\subref{fig:throughput-hot-write-5} the
comparison shows that \atomicrmiii{} significantly and consistently outperforms
Atomic RMI and all remaining frameworks, with the exception of HyFlow2.
Specifically, \atomicrmiii{} achieves at least a 47\% better throughput over
Atomic RMI due to the introduced optimizations. The impact of read-only
optimizations is visible in the 90\% read scenario, where \atomicrmiii{}
achieves up to a 201\% advantage in throughput.  Furthermore, the write
optimizations give \atomicrmiii{} a performance boost of up to 167\% over
Atomic RMI in the 90\% write scenario. In a more balanced scenario
optimizations can be applied less often, leading to a slightly lower
performance improvement of up to 72\%. Note, that Atomic RMI's performance does
not change with respect to the differences in workloads among scenarios, since
Atomic RMI is agnostic of operation types.
HyFlow2 and \atomicrmiii{} perform similarly in read dominated and balanced
scenarios, with HyFlow2 outperforming \atomicrmiii{} by up to 10\% in a 16-node system,
and \atomicrmiii{} outperforming HyFlow2 by as much in a 4 node system. The
similarities in performance stem from special handling of read-only variables
in both systems. However, in a write dominated scenario, \atomicrmiii{} has a 77\%
percent throughput advantage, which we again attribute to extensive
write-oriented optimizations employed in OptSVA-CF.

The 10-array scenarios in
\rfig{fig:throughput-hot-read-5}--\subref{fig:throughput-hot-write-5} yield
similar results, but here, \atomicrmiii{} manages to consistently outperform HyFlow2,
as well as other evaluated frameworks. 
This is because transactions have more objects to randomly select from,
transactions tend to contain shorter subsequences of operations on the same
objects, which allows \atomicrmiii{} to release more objects earlier. 

\rfig{fig:throughput-hotmild} shows changes in throughput as above, but with
longer transactions, that perform mild array accesses in addition to hot array
accesses. Hence each transaction performs 10 operations on the hot array and 10
operations on the mild array, in the same read-to-write ratios. Since
accesses on mild arrays never lead to conflicts, the average contention is much
lower in this scenario than the previous.
Because of this, throughput increases for each framework. \atomicrmiii{} performs
similarly to HyFlow2 in the balanced scenario (up to 2\% reduction or 8\%
improvement), slightly better in the read dominated (8--19\% improvement), and
significantly better in the write dominated scenario (64--76\%). 
Both HyFlow2 and \atomicrmiii{} perform significantly better than all other frameworks,
including Atomic RMI.
The results are similar to those in the previous scenario, but show that
\atomicrmiii{}'s advantage decreases in lower contention, which we attribute to the
overhead introduced by the instrumentation and asynchronous execution.

The abort rates of \atomicrmiii{} and Atomic RMI remain at 0\% throughout the
evaluation, while 60--89\% of HyFlow2 transactions abort and retry at least
once due to conflicts, depending on the scenario (see \rfig{fig:aborts2}). 
This means, that irrevocable operations are likely to be aborted and
re-executed.
On the other hand, \atomicrmiii{} manages to rival the efficiency of an
optimistic TM system while bypassing problems with irrevocable operations.

\begin{figure}
{\begin{tabularx}{\linewidth}{Xcccccc} \hline
Scenario, clients: & 64 & 128 & 256 & 512 & 768 & 1024 \\ \hline%
9$\div$1 ratio     & 66 & 74  & 79  & 86  & 84  & 89   \\
5$\div$5 ratio     & 60 & 70  & 75  & 83  & 87  & 87   \\
1$\div$9 ratio     & 66 & 74  & 79  & 86  & 84  & 89   \\
\hline
\end{tabularx}
}
\caption{\label{fig:aborts2} HyFlow2's abort rate for \rfig{fig:scalability}.}
\end{figure}

Throughout we see that \atomicrmiii{} significantly outperforms Atomic RMI and
other lock-based distributed concurrency control mechanisms, and performs
similarly to or better than a state-of-the-art optimistic distributed TM, all
without the need to use aborts and, thus, without complicating irrevocable
operation executions, and while employing the reflection-based mechanisms that
allow to use CF model. We also see that \atomicrmiii{} performs best in
read-dominated scenarios, but become really competitive in write-dominated
scenarios, where the buffering- and asynchrony-related write-oriented
optimizations make a real difference to throughput.

\section{Related Work}
\label{sec:rw}

Several distributed TM systems were proposed (see e.g.,
\cite{BAC08,CRCR09,KAJ+08,%
WKK12}). Most of them replicate a
non-distributed TM on many nodes and guarantee that replicas are consistent.
Their programming model is different from our distributed transactions. Other
systems extend non-distributed TMs with a communication layer, e.g., DiSTM
\cite{KAJ+08} extends \cite{CRCR09} with distributed coherence protocols.
The others include HyFlow2 \cite{TRP13}, a distributed TM operating in the
data flow model that is covered in \rsec{sec:evaluation}. HyFlow \cite{SR11} is
an earlier version of HyFlow2, implemented in Java on top of Aleph and DeuceSTM
and included control flow and data flow concurrency control algorithms,
including TFA \cite{SR12} and DTL2 (a distributed version of TL2 \cite{DSS06}).
HyFlow was compared with HyFlow2 in \cite{TRP13} and was shown to perform worse
then its successor. 

Distributed transactions are successfully used where requirements for strong
consistency meet wide-area distribution, e.g., in Google's Percolator
\cite{PD10} and Spanner \cite{COR12}. 
Percolator supports multi-row, ACID-compliant, pessimistic database
transactions that guarantee snapshot isolation. This is a much weaker guarantee
than expected from TM systems. Another drawback in comparison to DTM is
that writes must follow reads. 
Spanner provides semi-relational replicated tables with general purpose
distributed transactions. It uses real-time clocks and Paxos to guarantee
consistent reads. Spanner requires some \emph{a priori} information about
access sets and defers commitment like \atomicrmiii{}, but aborts on conflict.
Irrevocable operations are banned in Spanner.
Spanner transactions provide snapshot isolation and external consistency (akin
to real-time order), much weaker properties than considered sufficient in DTM.

A pessimistic TM (but not DTM) is proposed in \cite{MS12}, where read-only
transactions execute in parallel, but transactions that update are synchronized
using a global lock to execute one-at-a-time. This idea was improved upon in
Pessimistic Lock Elision (PLE) \cite{AMS12}, where a number of optimizations
were introduced, including encounter-time synchronization, rather than
commit-time. However, the authors show that sequential execution of update
transactions yields a performance penalty. In contrast, the algorithm proposed
here maintains a high level of parallelism regardless of updates. In
particular, the entire transaction need not be read-only for a variable that is
read-only to be read-optimized.

In \cite{MS12} the authors propose pessimistic non-distributed TM that runs
transactions sequentially (as in \cite{WSA08}) but allows parallel read-only
transactions.  Operations are synchronized by delaying writes of the write-set
location (with busy waiting). 
This is done using version numbers of transactions.  In contrast,
\atomicrmiii{} uses object versions for similar purposes, which enables early
release.  However, direct comparison is difficult, because \cite{MS12} aims at
non-distributed environments with fast access, while \atomicrmiii{} assumes
network communication with overheads.

SemanticTM \cite{DFK14} is another pessimistic (non-distributed) TM. Rather
than using versioning or blocking, transactions are scheduled and place their
operations in bulk into a producer-consumer queues attached to variables. The
instructions are then executed by a pool of non-blocking executor threads that
use statically derived access sets and dependencies between operations to
ensure the right order of execution. The scheduler ensures that all operations
of one transaction are executed 
consistently and in the right order. The transactions cannot abort, either
forcibly nor by manual override, but any operations can have redundant
executions (causing problems with irrevocable operations).  On the other hand, 
SemanticTM
is wait-free, whereas OptSVA-CF is only deadlock-free.

\section{Conclusions}
\label{sec:conclusions}

The paper introduced OptSVA-CF, a new pessimistic TM concurrency control
algorithm for use with the CF model with complex objects. The algorithm is
based on supremum versioning, which allows manual aborts (although does not
abort transactions due to conflicts), and implements a number of optimizations
based on OptSVA, aiming at better parallel execution of concurrent transactions.
Generalization from variables to complex objects requires categorization of
operations into three discrete groups: reads, aborts, and updates. It also
requires that two types of buffers are used, a copy buffer and a log buffer,
the latter of which allows pure initial writes to execute without
synchronization, just like in the variable model. OptSVA-CF is implemented as
\atomicrmiii{}, which performs better than its predecessor: Atomic RMI with
SVA, as well as a number of lock-based synchronization mechanisms. It can also
outperform HyFlow2, a state-of-the-art optimistic DF DTM, and does so without
aborting (i.e. without problems with irrevocable operations).
Given this, we show that a pessimistic system can be as well-performing as an
optimistic one, and introduce a CF DTM with competitive performance that was
lacking.


\bibliographystyle{abbrv}
\bibliography{top-arxiv}


\end{document}